\documentclass[11pt,a4paper]{article}
\usepackage{jheppub}

\usepackage{dsfont}

\usepackage{makecell} 

\usepackage[utf8]{inputenc}

\usepackage{mathtools}
\usepackage{tikz}

\usepackage{todonotes}
\numberwithin{equation}{section}

\def\ket#1{|{#1}\rangle}
\def\bra#1{\langle{#1}|}

\DeclareMathOperator{\Tr}{Tr}

\title{Simulating matrix models with tensor networks}
\author[a]{Enrico M. Brehm}
\author[a]{, Yibin Guo}
\author[b,a]{, Karl Jansen}
\author[c,d,e,f]{, and Enrico Rinaldi}
\affiliation[a]{CQTA, Deutsches Elektronen-Synchrotron DESY,\\Platanenallee 6, 15738 Zeuthen, Germany}
\affiliation[b]{Computation-Based Science and Technology Research Center, The Cyprus Institute, \\20 Kavafi Street, 2121 Nicosia, Cyprus}
\affiliation[c]{Quantinuum K.K., Otemachi Financial City Grand Cube 3F\\ 1-9-2 Otemachi, Chiyoda-ku, Tokyo, Japan}
\affiliation[d]{Interdisciplinary Theoretical and Mathematical Sciences (iTHEMS) Program, RIKEN, \\Wako, Saitama 351-0198, Japan}
\affiliation[e]{Center for Quantum Computing (RQC), RIKEN, \\Wako, Saitama 351-0198, Japan}
\affiliation[f]{Theoretical Quantum Physics Laboratory, Cluster of Pioneering Research, RIKEN\\ Wako, Saitama 351-0198, Japan}
\emailAdd{enrico.brehm@desy.de}
\emailAdd{yibin.guo@desy.de}
\emailAdd{karl.jansen@desy.de}
\emailAdd{enrico.rinaldi@quantinuum.com}

\abstract{Matrix models, as quantum mechanical systems without explicit spatial dependence, provide valuable insights into higher-dimensional gauge and gravitational theories, especially within the framework of string theory, where they can describe quantum black holes via the holographic principle. Simulating these models allows for exploration of their kinematic and dynamic properties, particularly in parameter regimes that are analytically intractable. In this study, we examine the potential of tensor network techniques for such simulations. Specifically, we construct ground states as matrix product states and analyse features such as their entanglement structure.}

\keywords{matrix model, tensor network, string theory, simulation, BFSS model, BMN model}
\begin{document}
\maketitle

\section{Introduction}

Matrix models are quantum mechanical models whose degrees of freedom are organised in matrices. 
Although they do not directly depend on space, they can offer a way to study theories in higher dimensions, for example supersymmetric gauge theories. 
In particular, they can provide a non-perturbative formulation of certain quantum field theories and string theories, allowing for a more comprehensive understanding of these theories and their dynamics.
See Ref.~\cite{Maldacena:2023acv} for a recent review.

The gauge/gravity duality~\cite{Maldacena:1997re}, which establishes a deep connection between problems in quantum and gravitational physics, has further motivated the study of matrix models. 
This duality maps certain strongly-coupled quantum field theories to weakly-curved gravitational theories in higher dimensions. 
Via this duality, the study of matrix models can also discover insights into gravity in higher dimensions and may, for example, reveal more about features of black holes.

Simulating higher-dimensional quantum field theories and string theories poses significant challenges. 
Discretising these systems on a spatial lattice is a somewhat artificial construction, which may make it difficult to capture the full underlying physics. 
Leveraging dualities and constructions from string theory can offer a more physical realisation of higher-dimensional quantum field theories, potentially leading to more efficient and accurate simulations. 
Particularly, matrix model constructions are free from any need of discretising spacetime and may open a path to simulate aspects of higher-dimensional quantum field theories and string theory.

The study of matrix models has a rich history, dating back to the early days of string theory and quantum gravity research. 
Notable examples include the IKKT model~\cite{Ishibashi:1996xs}, which provides a non-perturbative definition of type IIB superstring theory, and the BFSS model~\cite{Banks:1996vh,Seiberg:1997ad}, which describes M-theory in the infinite momentum frame and will be discussed in Sec.~\ref{sec:BFSS}. 
These models have massively advanced our understanding of string theory and quantum gravitational theories in general.

The complexity of matrix models, particularly in the large-$N$ limit, where $N$ is the size of the matrices, necessitates the development of powerful computational techniques. 
Traditional methods, such as perturbative expansions in large $N$, often fall short in capturing the full non-perturbative dynamics of these systems. 
This has led to more and more interest in applying advanced numerical methods to study matrix models, ranging from Monte Carlo simulations (see, e.g., Ref.~\cite{Anagnostopoulos:2007fw,Berkowitz:2016jlq,Bergner:2021goh,Pateloudis:2022ijr}), to bootstrap methods (see, e.g.,Ref.~\cite{Han:2020bkb,Lin:2023owt}), to machine learning algorithms (see, e.g., Ref.~\cite{Han:2019wue,Rinaldi:2021jbg,Bodendorfer:2024egw}), and, recently, to quantum computing approaches (see, e.g., Ref.~\cite{Gharibyan:2020bab,Rinaldi:2021jbg}).

In this work we want to focus on tensor network methods as a versatile framework for approximating and simulating matrix models. 
Tensor networks, originally developed in the context of condensed matter physics to study strongly correlated quantum systems~\cite{Orus:2018dya}, have emerged as a promising tool for tackling the computational challenges in other areas of physics~\cite{Magnifico:2024eiy}. 
They are also a promising tool to tackle the challenges posed by matrix models: the ability of tensor networks to efficiently represent and manipulate high-dimensional quantum states makes them well-suited for exploring the large Hilbert spaces characteristic of matrix models at larger $N$.

Our analysis aims to shed light on the strengths and limitations of tensor network method when applied to matrix models. 
We will examine aspects such as computational efficiency, scalability to larger matrix sizes, and the ability to capture specific physical effects and observables. 
By doing so, we hope to provide valuable insights into the most effective strategies for simulating matrix models, ultimately contributing to the understanding of quantum field theories, string theory, and the nature of quantum gravity.
Our results provide a comprehensive overview of the convergence of tensor network methods as matrix models grow in size and complexity, both in terms of the size $N$ and the number of matrices, as well as the truncation of the local Hilbert space.
In practice, we provide a guide to the approximation error of these numerical methods as a function of the bond dimension in the entire complexity landscape of bosonic and fermionic matrix models.

The remainder of the paper is organised as follows.
In Sec.~\ref{sec:matrixModels}, we provide an overview of two important matrix models, the BFSS and BMN models, and their connection to string theory.
We then show how to simulate matrix models in the Hamiltonian formulation in Sec.~\ref{sec:simulation} and discuss how tensor networks methods are applied to the task of finding the ground state.
Our results across a range of models are presented in Sec.~\ref{sec:numerical}, including an analysis of the entanglement structure of the ground states and of the computational efficiency of the tensor network simulations. The numerical results look very promising. Our results align with previous findings from simulations employing machine learning and quantum computing approaches, as e.g. presented in Ref.~\cite{Han:2019wue,Rinaldi:2021jbg,Bodendorfer:2024egw} and Ref.~\cite{Gharibyan:2020bab,Rinaldi:2021jbg}. However, our analysis of the computational cost for finding ground states indicate a polynomial scaling with the number of bosonic degrees of freedom, enabling us to access matrix sizes and quantities beyond the reach of these methods. This demonstrates that tensor networks are a promising and scalable tool for simulating matrix models involving many large matrices.

Finally, we conclude in Sec.~\ref{sec:conclusion} with a summary of our findings and an outlook on future research directions.
More technical details are provided in the appendices.

\section{Matrix models and the connection to String/M-Theory}\label{sec:matrixModels}

As teased already, matrix models play a crucial role in string theory, offering unique insights into non-perturbative regimes that are often challenging to explore through conventional methods. They serve as powerful tools for probing the fundamental structure of string theory and its various manifestations. In this section, we present a short overview of two important matrix models and their diverse contexts within string theory without assuming too much background in the latter. The rich duality structure in string theory allows these models to emerge in remarkably varying scenarios, showcasing their fundamental importance and versatility. By examining these models, we can gain a deeper understanding of the interconnected nature of different formulations of string theory and potentially uncover new ways for exploring quantum gravity.

\subsection{The BFSS matrix model}\label{sec:BFSS}

\begin{figure}
    \centering
    \includegraphics[width=0.5\linewidth]{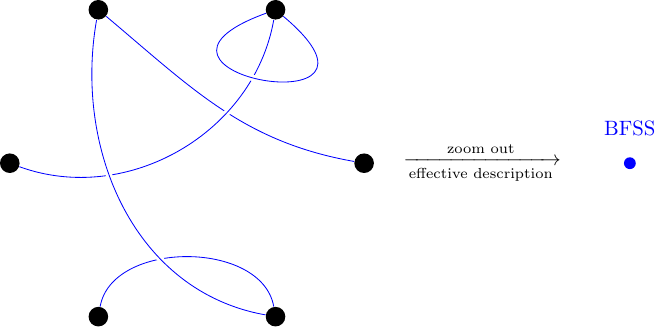}
    \caption{Six D0-branes (represented by the black dots) with some strings (in blue) between them. Zooming out a potential low energy  effective description of the dynamics of these six branes is governed by the BFSS matrix model of $6\times6$ matrices. }
    \label{fig:BFSS}
\end{figure}

Among the first matrix models that appeared in the context of string theory was the so-called BFSS matrix model \cite{Banks:1996vh,Seiberg:1997ad}. It is the effective description of the dynamics of interacting \textit{D0-branes}\footnote{A Dirichlet membrane, or short D-brane, is an extended object upon which open strings can end with Dirichlet boundary conditions. Usually they are categorised by their spatial dimension which is indicated by a number after the D. Hence, a D0-brane is a spatial point, a D1-brane is a spatial line, a D2-brane a spatial surface, and so on.}. For six D0-branes the rough picture is indicated in Fig.~\ref{fig:BFSS}. Particularly for a large number $N$ of branes it encodes specific features of \textit{M-theory}.\footnote{M-theory is the conjectured unified description of all consistent versions of supersymmetric string theory with 11 dimensional supergravity as its low energy limit. The idea of a unifying parent theory originates from the discovery of an extended web of dualities between all the known string theories.}

The BFSS model also arises as the Kaluza-Klein compactification\footnote{Kaluza-Klein compactification is a theoretical approach where extra spatial dimensions are "curled up" so tightly, typically modeled as a small circle or other compact geometric shape, that they are too small to detect.} of $10D$ super Yang-Mills theory to (1+0) dimensions \cite{Polchinski:1996na,Ishibashi:1996xs} and as a description of so-called M2-branes\footnote{M2-branes are solutions to 11-dimensional supergravity. They are extended objects of spatial dimension 2 and can to some extent be regarded as an analogue to the idea of fundamental strings from standard string theory with one more spatial dimension.} \cite{Nicolai:1998ic,Dasgupta:2002iy}.

All these examples and frameworks in which this specific matrix model appears suggest that studying it can, in principle, provide valuable insights into advanced theories that may describe higher dimensional quantum systems, including quantum gravity. In fact, the BFSS model has been used to compute graviton scattering amplitudes \cite{Becker:1997wh,Becker:1997xw,Helling:1999js,Herderschee:2023pza}, test the soft graviton theorem \cite{Miller:2022fvc,Tropper:2023fjr,Herderschee:2023bnc}, compute features of black holes \cite{Banks:1997hz,Banks:1997tn,Klebanov:1997kv,Halyo:1997wj,Horowitz:1997fr,Kabat:1997im,Hyakutake:2018bht,Du:2018dmi}, and has been connected to lattice gauge theory \cite{Filev:2015hia,Hanada:2016jok,Bergner:2021goh}. 

Explicitly, the BFSS model is a quantum mechanical model of 10 bosonic $N\times N$ Hermitian matrices $A_t(\tau)$ and $X_I(\tau)$, $I=1,\dots,9$, together with a 16-component spinor $\psi_{\alpha,\beta}(\tau)$, with each element an $N\times N$ fermionic matrix, specified by the action 
\begin{equation}\label{eq:BFSSaction}
    S_\text{BFSS} = \frac{1}{g^2}\int d\tau\,  \mathrm{Tr}\left[\frac{1}{2}(D_t X_I)^2 +\frac{1}{4} \left[X_I,X_J\right]^2  + \frac{i}{2}\psi^\dagger D_t\psi - \frac{1}{2}\psi^\dagger\gamma^I\left[X_I,\psi\right] \right] \,,
\end{equation}
where $D_t(\cdot) = \frac{\partial}{\partial\tau}(\cdot) - i g [A_t,(\cdot)]$ is a covariant derivative, $\gamma^I$ are ($16\times16$) representation of the Clifford algebra, and free indices are summed over. 

The model clearly captures some interesting features of quantum gravity and in particular M-theory. However, it is also clear that it cannot be the full answer. It, for example, needs the definition of an asymptotic Minkowski background and a particular scaling of the coupling over length of strings. 
This means that it can only capture a particular corner of the full M-theory. 
Another problem is the question about the existence of a sensible ground state of the BFSS model \cite{deWit:1988xki}. It has a continuous spectrum and, in particular, no gap above the vacuum.  This, for example, leads to the conclusion that membranes in Minkowski space are (perturbatively) unstable. Attempts to solve this issue were presented regularly over the years, see, e.g., Ref.~\cite{Boulton:2015mha,Boulton:2018yol} and references therein. 

\subsection{The BMN matrix model}\label{sec:BMN}

Some of the latter problems can be resolved by introducing a mass deformation of the BFSS matrix model, which is then dubbed BMN matrix model \cite{Berenstein:2002jq,Dasgupta:2002hx}. 
In the string theory context it arises in the description of M2-branes in so-called pp-wave spacetimes.\footnote{Pp-wave spacetimes are families of exact solutions of Einstein's field equation. It is basically Minkowski space filled with plane waves with parallel propagation \cite{Ehlers:1962zz}.} Via dualities, it is also connected to bound states of D2-D0 brane configurations in type IIA string theory \cite{Lin:2004kw} or black M-brane\footnote{Black brane solutions are generalisations of black holes solutions. In particular in higher dimensions one can not only take the limiting configuration of a point mass gravitational source, but also those supported on a line, a surface, or a higher dimensional space. The respective solutions to Einstein's field equations are called black branes. See, e.g., Ref.~\cite{Duff:1993ye}. } configurations in different compactification backgrounds \cite{Berenstein:2002jq}. One can also realise various supersymmetric quantum field theories within the BMN matrix model by taking suitable vacua \cite{Berenstein:2002jq}. This includes for example 3d super Yang-Mills theory (SYM), 6d superconformal field theory, and 4d SYM. Also, due to the connection to black branes the thermal behaviour \cite{Shin:2005iy,Asplund:2011qj,Brady:2013opa,Costa:2014wya,Pramodh:2014jha}, phase transitions  and confinement \cite{Asano:2018nol,Asano:2020yry}, chaos \cite{Axenides:2017nwp}, and entanglement \cite{Gray:2020zov} of the BMN matrix model were discussed over the years. 

Explicitly, the BMN matrix model for some given $N$ has the same field content as the BFSS matrix model. However, it is now convenient to divide the scalars into $X_i$ with $i=1,2,3$ and $X_a$ with $a=4,\dots,9$. Then the BMN action can be written as \cite{Gharibyan:2020bab}
\begin{equation}
\begin{split}
    S_\text{BMN} = \frac{1}{g^2}\int d\tau\,  \mathrm{Tr}\Big[&\frac{1}{2}(D_t X_I)^2 +\frac{1}{4} \left[X_I,X_J\right]^2  -\frac{\mu^2}{18} X_i^2 - \frac{\mu^2}{72} X_a^2 - \frac{i\mu}{6}\epsilon^{ijk}X_iX_jX_k\\
    & +\frac{i}{2}\psi^\dagger D_t\psi - \frac{1}{2}\psi^\dagger\gamma^I\left[X_I,\psi\right]  -\frac{i\mu}{8} \psi^\dagger \gamma^{123}\psi\Big] \,, 
\end{split}
\end{equation}
where, in addition to the objects already defined for Eq.~\eqref{eq:BFSSaction}, we have the mass deformation parameter $\mu$, the totally antisymmetric tensor $\epsilon^{ijk}$ in the so-called Myers term and the notation $\gamma^{ij\dots} = \gamma^i\gamma^j \cdots$. The model exhibits a $U(N)$ redundancy, has sixteen supercharges, a global $SO(3)$ symmetry rotating the 3 scalar matrices $X_i$, and an $SO(6)$ symmetry for the remaining 6 scalar matrices $X_a$. 

\subsection{Simpler toy models}\label{sec:toyModel}

In this work, we aim to conduct numerical studies of matrix models. However, the models discussed above, such as the BMN matrix model, possess a rich field content that renders a full simulation computationally challenging. To make our analysis more tractable, we focus on simplified models with reduced field content while preserving essential features such as $SU(N)$ invariance and supersymmetry.

Fortunately, a comprehensive classification of massive super Yang-Mills quantum mechanics exists in Ref.~\cite{Kim:2006wg}, providing a spectrum of models that can be viewed as analogues of the BFSS or BMN matrix models. There, these models are derived through dimensional reduction and mass deformation of supersymmetric Yang-Mills theories in 6D, 4D, and 3D.

For example, we can consider a matrix model originating from the dimensional reduction of the unique minimal $\mathcal{N}=1$ SU$(N)$ super Yang-Mills theory in three dimensions. This reduction yields an $\mathcal{N}=2$ supersymmetric matrix model comprising two $\mathfrak{su}(N)$ matrices, $X_i$ ($i=1,2$), governed by the Lagrangian:
\begin{equation}
    \mathcal{L}^{\mathcal{N} = 2}_0 = \Tr\left[\sum_{i=1}^2\frac{1}{2} D_t X_i D_t X_i + \frac{1}{2} \left[X_1,X_2\right]^2 - \frac{i}{2} \bar{\psi}\Gamma^t D_t \psi - \sum_{i=1}^2\frac{1}{2} \bar{\psi} \Gamma^i\left[X_i,\psi\right]\right]\,,
\end{equation}
where $\psi = (\xi,i\xi^\dagger)$ with $\xi$ being fermionic $N\times N$ matrices, and $\Gamma^a$ are the Gamma matrices $\Gamma^t = i\sigma_3$, $\Gamma^1 = \sigma_1$, and $\Gamma^2 = \sigma_2$, and $D_t$ is again the covariant derivative. Introducing an interaction constant $g$ and a mass deformation without breaking $\mathcal{N}=2$ SUSY gives 
\begin{equation}\label{eq:toymodel}
\begin{split}
    \mathcal{L}^{\mathcal{N} = 2}_{g,m} =~~&\Tr\left[\frac{1}{2} \sum_{i=1}^2 D_t X_i D_t X_i - \frac{m^2}{2} \left(X_1^2 + X_2^2\right) + \frac{g^2}{2} \left[X_1,X_2\right]^2 \right] \\
     - &\Tr\left[ i \frac{1}{2} \bar{\psi}\Gamma^t D_t \psi - i \frac{3m}{4} \bar{\psi}\psi + \frac{g}{2} \bar{\psi} \Gamma^1\left[X_1,\psi\right] + \frac{g}{2} \bar{\psi} \Gamma^2\left[X_2,\psi\right]  \right]\,.
\end{split}
\end{equation}

\subsubsection{Bosonic matrix models}

To make our analysis even more tractable we may also drop the necessity of supersymmetry and fermionic degrees of freedom. As we will see in the upcoming sections, most of the numerical challenge originates from the bosons, on which we want to focus most of our attention. The model we will consider is the straightforward extension of the bosonic part of Eq.~\eqref{eq:toymodel} from 2 to $D$ bosonic $N\times N$ matrices:
\begin{equation}\label{eq:BosToyModel}
\begin{split}
    \mathcal{L}_{g,m} = \Tr\left[ \sum_{i=1}^D \left(\frac{1}{2} D_t X_i D_t X_i - \frac{m^2}{2} X_i^2\right) + \frac{g^2}{2} \sum_{i<j}\left[X_i,X_j\right]^2 \right]\,.
\end{split}
\end{equation}
Note, that in the present paper for simplicity we also omit the interaction term that is proportional to $\epsilon^{ijk} X_iX_jX_k$ that usually appears for the models constructed in \cite{Kim:2006wg} with more than two matrices and also in BMN model. 

\section{Simulation of matrix models}\label{sec:simulation}

The aforementioned matrix models can offer the opportunity to study quantum gravitational or gauge theories in higher dimensions in a more tractable manner. 
They are, in particular, quantum mechanical models and have no dependence on spatial coordinates.
Hence, when trying to simulate them there is no need of a spatial lattice formulation. 
This can, for example, save us from issues with symmetries broken on the lattice that need to be recovered in the continuum limit. 
Chiral symmetries and supersymmetry are examples that cannot be preserved on the lattice and are hard to recover. 
In quantum mechanics, although one may still need a regularisation that breaks symmetries, very often this is not a problem due to the lack of ultraviolet divergences~\cite{Gharibyan:2020bab}. 
We will see that the degrees of freedom of the hermitian matrices can be expressed as a collection of interacting harmonic oscillators, which allows for a straightforward truncation scheme in the respective Hilbert space.

Another advantage in the quantum mechanical context of matrix models is that there is no dynamical gauge field. 
Symmetries are always ``global'' and singlet constraints can be imposed directly on the states~\cite{Rinaldi:2021jbg}.  

\subsection{Hamiltonian formulation}\label{sec:Hamiltonian}

In this work we are aiming to simulate matrix models with tensor network states, which is usually done in the Hamiltonian formulation. 
The fields introduced in the previous action formulations get promoted to operators acting on the Hilbert space of the matrix model.
We first want to focus on the rather simple matrix model specified by Eq.~\eqref{eq:toymodel}. 
Its Hamiltonian is given by
\begin{align}
    H  ~~=\,& H_X + H_{\psi} - (N^2 - 1) m\,,\\
    H_X  =\,& \Tr\left( \frac{1}{2} \left(P_1^2 + P_2^2\right)  + \frac{m^2}{2} \left(X_1^2 + X_2^2\right)- \frac{g^2}{2} \left[X_1,X_2\right]^2   \right) \,,\\
    H_{\psi}\,=\,& \Tr\left(  -i \frac{3m}{4} \bar{\psi} \psi + \frac{g}{2} \bar{\psi} \Gamma^1\left[X_1,\psi\right] + \frac{g}{2} \bar{\psi} \Gamma^2\left[X_2,\psi\right]\right) \\
    =& \Tr\left(\frac{3m}{2} \xi^\dagger \xi - \frac{g}{2} \xi \left[X_1 + i X_2,\xi \right] - \frac{g}{2} \xi^\dagger \left[X_1 - i X_2,\xi^\dagger \right]\right)\,,
\end{align}
where $P_i$ are the conjugate variables to $X_i$. 
The shift $-(N^2-1)m$ is such that the vacuum energy is vanishing. 
Expanding the hermitian operators in $\mathfrak{su}(N)$ generators and rewriting the bosonic degrees of freedom in terms of lowering and raising operators (see App.~\ref{app:su(N)expansion}) we can write the Hamiltonian as
\begin{align}\label{eq:HamiltonianBos}
    H_X =& \,\sum_\alpha (n_{1\alpha} + n_{2\alpha} + 1) m + \frac{g^2}{2} \sum_{\alpha,\beta,\gamma,\rho,\sigma} f_{\alpha\beta\sigma} f_{\gamma\rho\sigma} \,X_{1\alpha} X_{2\beta} X_{1\gamma} X_{2\rho}\,,\\
    \label{eq:HamiltonianFerm}
    H_{\psi}\, = &\, \sum_\alpha\frac{3m}{2} \xi^\dagger_\alpha\xi_\alpha + \frac{g}{2}\sum_{\alpha,\beta,\gamma} f_{\alpha\beta\gamma} \, \xi_\alpha (X_{2\beta} - i X_{1\beta}) \xi_\gamma -  f_{\alpha\beta\gamma} \, \xi^\dagger_\alpha (X_{2\beta} + i X_{1\beta}) \xi^\dagger_\gamma \,,
\end{align}
where the Greek letters run from 1 to $N^2-1$, the dimension of $\mathfrak{su}(N)$, and $f_{\alpha\beta\gamma}$ are the structure constants whose explicit form are given in App.~\ref{app:structureConst}. 

The state space of bosonic excitations is created from the space of $2(N^2-1)$ free quantum harmonic oscillators, which is also called Fock space (see App.~\ref{app:statespace}). 
The Fock space is infinite dimensional whereas in a simulation we can only handle finite-dimensional Hilbert spaces. 
We can tackle the problem by truncating the local Fock spaces and restrict the bosonic excitation levels of the various oscillators. 
We call the truncation cutoff $\Lambda$, i.e. every boson $(i,\alpha)$ can at most be excited $\Lambda$ times (see Eq.~\eqref{eq:truncation}).  

For the purely bosonic toy model with Lagrangian given in Eq.~\eqref{eq:BosToyModel} the Hamiltonian is given by
\begin{align}
    H &= \Tr\left( \frac{1}{2} \sum_{i=1}^D \left(P_i^2 + m^2 X_i^2\right) - \frac{g^2}{2} \sum_{i<j}\left[X_i,X_j\right]^2 \right)\\
     &= \sum_{i,\alpha} m \left(n_{i\alpha} + \frac{1}{2}\right) + \frac{g^2}{2} \sum_{i<j} \sum_{\alpha,\beta,\gamma,\rho,\sigma} f_{\alpha\beta\sigma} f_{\gamma\rho\sigma} \,X_{i\alpha} X_{j\beta} X_{i\gamma} X_{j\rho} \,.
\end{align}

\subsection{Symmetries of the supersymmetric minimal BMN-like matrix model} \label{sec:symmetries}

The model exhibits an SU$(N)$ symmetry that originates from the respective gauge symmetry in three dimensions.
We can regard it  as a gauge symmetry for the matrix model as well. 
When using the gauge $A_t = 0$, one has to impose Gauß's law \cite{Gharibyan:2020bab}
\begin{equation}
    \mathcal{G} = i \left(2\sum_{i=1}^2 \left[ D_t X_i,X_i \right] + \left[ \bar\psi,\psi \right] \right) = 0 
\end{equation}
which, in particular, restricts all states to be singlets under the SU$(N)$ symmetry. 
With the expansion from Eq.~\eqref{eq:SUNexpansion} we can write
\begin{equation}
    G_\alpha = i \sum_{\beta\gamma} f_{\alpha\beta\gamma} \left(\sum_{i} a_{i\beta}^\dagger a_{i\gamma} + \xi^\dagger_\beta\xi_\gamma\right)\,
\end{equation}
which act as generators of the symmetry on the state space and commute with the Hamiltonian of the matrix model. 
A gauge invariant state $\ket{s}$ has to fulfil $G_\alpha \ket{s} = 0$ for all $\alpha$. 
Then, these states are dubbed physical. In this work we will mostly regard the SU$(N)$ symmetry as a gauge and will construct singlet states.

However, in contrast to gauge theories in higher dimension, in a purely quantum mechanical model it is possible to regard the SU$(N)$ symmetry as a global symmetry. 
Then in an ungauged model also non-singlet states are in principle allowed. 
This was e.g. discussed in Refs.~\cite{Maldacena:2018vsr,Berkowitz:2018qhn} for the BFSS model.
There, the dual gravitational picture to the ungauged matrix model includes particular non-local objects called Wilson lines. 
However, an extended study where we consider also non-singlet states will be left for the future. 

The model also exhibits supersymmetry which relates its bosonic and fermionic degrees of freedom.
Its infinitesimal action on the matrices is given in Ref.~\cite[eq. (6.4)]{Kim:2006wg} and the two SUSY generators $Q$ and $Q^\dagger$ are given by \cite[eq. (8)]{Rinaldi:2021jbg}
\begin{equation}
    Q = - \sum_\alpha \xi_\alpha \left(\left(P_{1\alpha} - iP_{2\alpha}\right) - i \,m \left(X_{1\alpha} - iX_{2\alpha}\right) \right) - i\frac{g}{\sqrt{2}} \sum_{\alpha,\beta,\gamma} f_{\alpha\beta\gamma} \xi_\alpha X_{1\beta} X_{2\gamma}\,.
\end{equation}
Next, there is a global SO$(2)$ symmetry that rotates the two bosonic matrices into each other, i.e. for a rotation angle $\theta$ we have
\begin{align}
    X_1 &\mapsto \cos(\theta) X_1 + \sin(\theta) X_2\,,\\
    X_2 &\mapsto \cos(\theta) X_2 - \sin(\theta) X_1\,,\quad\text{and}\\
    \xi &\mapsto e^{i\theta/2} \xi\,,
\end{align}
which is generated by the angular momentum operator
\begin{equation}
    L = \sum_\alpha  X_{2\alpha}P_{1\alpha} - X_{1\alpha} P_{2\alpha} - \frac{1}{2}\xi_\alpha^\dagger\xi_\alpha\,.
\end{equation}
Note that in case of more matrices, there are larger global symmetries rotating the matrices of equal mass into each other. In case of the bosonic toy model Eq.~\eqref{eq:BosToyModel} there is a global SO$(D)$ symmetry.  

The truncation of the boson Hilbert space, which is necessary for simulating the model, usually breaks all symmetry. Hence, the truncated symmetry generators do not commute with the truncated Hamiltonian, e.g. for the SU$(N)$ generators
\begin{equation}
    \left[H,G_\alpha\right] \neq 0\,
\end{equation}
and $H$ and $G$ do no longer share a common eigenbasis. However, the symmetry breaking is expected to vanish for states with energies much smaller than the cutoff scale $\Lambda$, i.e. 
\begin{equation}
    \bra{\psi} [H,G_\alpha] \ket{\psi} = \epsilon \ll 1\,,\quad \mathrm{for}~E_\psi \ll \Lambda\,.
\end{equation}
When constructing the spectrum one can enforce the Gauge constraint 
\begin{equation}
    G_\alpha \ket{\psi} = 0
\end{equation}
for example by adding a penalty term proportional to $G^2 := \sum_\alpha G_\alpha^2$ to the Hamiltonian, i.e.
\begin{equation}\label{eq:panalty}
    H' = H + p \,G^2\,.
\end{equation}
For large $p$, states with energies much lower than $p$ should then be gauge singlets. 
When we consider a cutoff $\Lambda$ then choosing $p\sim\Lambda$ should render the low-lying spectrum gauge invariant. 
In addition, for large enough $\Lambda$, $E'_\psi \approx E_\psi$ and the singlet condition should be preserved to good precision. 

\subsection{Tensor network simulation}\label{sec:tensorNetwork}

\subsubsection{Matrix product states and matrix product operators}

Tensor network states (TNS) have, in particular, become powerful mathematical and numerical tools to simulate the ground state and low-energy excited states of several quantum (many-body) systems~\cite{white1992density,Orus:2018dya,Banuls:2022vxp,Magnifico:2024eiy,perez2006matrix,schollwock2011density,bridgeman2017hand,paeckel2019time,hauschild2018efficient,xiang2023density,pirvu2010matrix,chan2016matrix}. The core idea is to describe a high-rank tensor with a network of lower-rank tensors, significantly reducing the computational complexity while preserving essential physical properties.
The structure of TNS usually respects the spatial structure of the physical system it represents. 
For example, matrix product states (MPS)~\cite{ostlund1995thermodynamic} and matrix product operators (MPO)~\cite{pirvu2010matrix,chan2016matrix} follow a one-dimensional structure and are particularly effective for representing states and operators of one-dimensional quantum systems. 
In App.~\ref{app:MPSandMPO} we give a short explanation of MPS and MPO. 
In this work we choose these tensor networks to simulate matrix models.  

\subsubsection{Density Matrix Renormalization Group algorithm}

In this paper, we employ the Density Matrix Renormalization Group (DMRG) algorithm to determine the MPS representing the ground state~\cite{white1992density,fishman2022itensor}. 
As a variational method, it minimizes the energy functional
\begin{equation}
    E(\{ A^{i} \}) = \frac{\langle \psi |H| \psi \rangle}{\langle \psi | \psi \rangle}
\end{equation}
by optimising two neighbouring $A^i$ at a time, repeatedly sweeping through the chain until convergence. Here, $A^i$ are the local 3-tensors that built the MPS of the state $\ket{\psi}$, see Eq.~\eqref{eq:MPS}.
With a diagrammatic notation as in Eq.~\eqref{eq:MPSgraph}, the energy functional can be obtained by contracting the MPS representing the current state and the MPO of the Hamiltonian. 
During the sweep, we need to solve a generalized eigenvalue problem to update the local tensors and reducing the energy. 
This step usually increases the bond dimension $D$. 
To maintain manageable computational costs a low-rank approximation might be necessary. 
This approximation is performed using a singular value decomposition (SVD) and keeping $D_\mathrm{max}$ basis vectors that correspond to the largest singular values. 
The discarded singular values contribute to the truncation error as
\begin{equation}
    TE = \frac{\sum_{m=dD_{\mathrm{max}}+1}^{dD} s_{m}^{2}}{\sum_{m=1}^{dD_{\mathrm{max}}} s_{m}^{2}},
\end{equation}
where $d$ is the local physical dimension of the tensors and the singular value $s_m$ are arranged with descending order. 
Here, $TE$ describes the quadratic norm distance between the approximated and original wave function. 

Denoting the bond dimension of the MPO as $\chi$, the two-site update algorithm for DMRG then has a computational cost of
\begin{equation}
   \mathcal{O}(2 L D_{\mathrm{max}}^{3} \chi d + 2 L D_{\mathrm{max}}^{2} \chi^{2} d^{2} + L D_{\mathrm{max}}^{3} \chi d^{2})
   \label{eq:MPScost}
\end{equation}
for each complete sweep. 
Selecting $D_\mathrm{max}$ requires balancing computational complexity with the accuracy of preserving key physical properties.

\subsubsection{Entanglement entropy and the entanglement spectrum}\label{sec:entanglement}

As a method inspired by quantum information theory, matrix product states (MPS) are particularly well-suited for determining entanglement properties, including entanglement entropy (EE) and the entanglement spectrum (ES). The EE provides an estimate of the quantum resources required to simulate the states. The ES, serving as a complement to the EE, can reveal additional information, such as the topological properties of the quantum state.

Let us start from the density matrix $\rho \equiv |\psi\rangle \langle\psi|$ corresponding to the MPS. Suppose the system is bi-partitioned into two subsystems, naming $A$ and $B$. After tracing out the degree of freedom of the subsystem $B$, the entanglement entropy $S_{A}$ is given by
\begin{equation}
    S_{A} = -\mathrm{Tr}(\rho_{A} \mathrm{log} \rho_{A}),
\end{equation}
where $\rho_{A}$ is the reduced density matrix of subsystem $A$. The entanglement Hamiltonian $\mathcal{H}_{E}$ is given by the relation \cite{Li:2008}
\begin{equation}
    \rho_{A} = \frac{e^{-\mathcal{H}_{E}}}{\mathrm{Tr}(e^{-\mathcal{H}_{E}})},
\end{equation}
and the entanglement spectrum $\zeta_{m}$ is given by the eigenvalues of $\mathcal{H}_{E}$. If a Schmidt decomposition between the system $A$ and $B$ is known, then both the entanglement entropy and the entanglement spectrum can be expressed in term of the respective coefficients. Explicitly, for a normalised MPS in canonical form (see App.~\ref{app:canonicalform}), the Schmidt coefficients are the eigenvalues $s_m$ of the matrix $C^{i}$ given in Eq.~\eqref{eq:MPSCoefCano} and entanglement entropy is
\begin{equation}
    S_{A} = -\sum_{m=1}^{D_{i}} s^{2}_{m} \mathrm{log} s^{2}_{m},
\end{equation}
and the entanglement spectrum is
\begin{equation}\label{eq:entEigenvalues}
    \zeta_{m} = -2\mathrm{log} s_{m}.
\end{equation}

\section{Numerical analysis of the models}\label{sec:numerical}

In this section, we present numerical results for matrix models introduced in Sec.~\ref{sec:toyModel}, demonstrating the feasibility of using MPS to simulate such models. We begin by discussing the strategy for selecting a proper mapping between the different particles and a one-dimensional lattice for the MPS simulation. As a proof of concept, we provided detailed results, including ground state energies and the computational cost for interacting bosonic SU$(2)$ matrices, with varying the bosonic cutoff $\Lambda$ and the number of matrices. To further validate the feasibility of this approach to other SU$(N)$ matrix models and supersymmetric models, we also present the results for the minimal BMN-like matrix model \ref{eq:toymodel}.

\subsection{Layout schemes of matrix models}

To implement the MPS simulation on the matrix models, the first step is to arrange the degrees of freedom of the models in a one-dimensional chain. After the arrangement, the Hamiltonian of the matrix models can be represented with an MPO. Different layout schemes results in MPOs with varying bond dimensions, which directly determine the computational cost according to Eq.~\eqref{eq:MPScost}. We conduct a preliminary study on the scaling of the maximal bond dimension of the MPO for various layout schemes in App.~\ref{app:layoutscheme}. Instead of conducting a throughout computational cost analysis that includes the scaling behaviour of both the MPS and MPO bond dimension, we only focus on the MPO bond dimension analysis. This approach provides a quick and effective way for identifying preferred layout schemes in terms of computational cost. We leave an extended but more resource heavy analysis for future studies. 

The observations shown in App.~\ref{app:layoutscheme} are a first hint that MPS and MPO are promising tools to simulate matrix models, particular for more than two matrices, as the computational cost contributed by the MPO seems to increases at most linearly with the number of bosonic fields.

\subsection{Results for SU(2) bosonic matrix models}

After mapping the SU$(2)$ bosonic matrix models to chains using the sequential layout scheme, we execute the DMRG algorithm to determine the ground states. The stopping criterion is based on the energy difference between consecutive sweeps, with a threshold set at $10^{-8}$. We do not include the penalty terms, as the gauge-invariant condition of the ground state should be automatically satisfied when the cutoff $\Lambda$ approaches infinity, as discussed at the end of Sec.~\ref{sec:symmetries}. 

\begin{figure}[htbp!]
    \centering
    \includegraphics[width=0.85\linewidth]{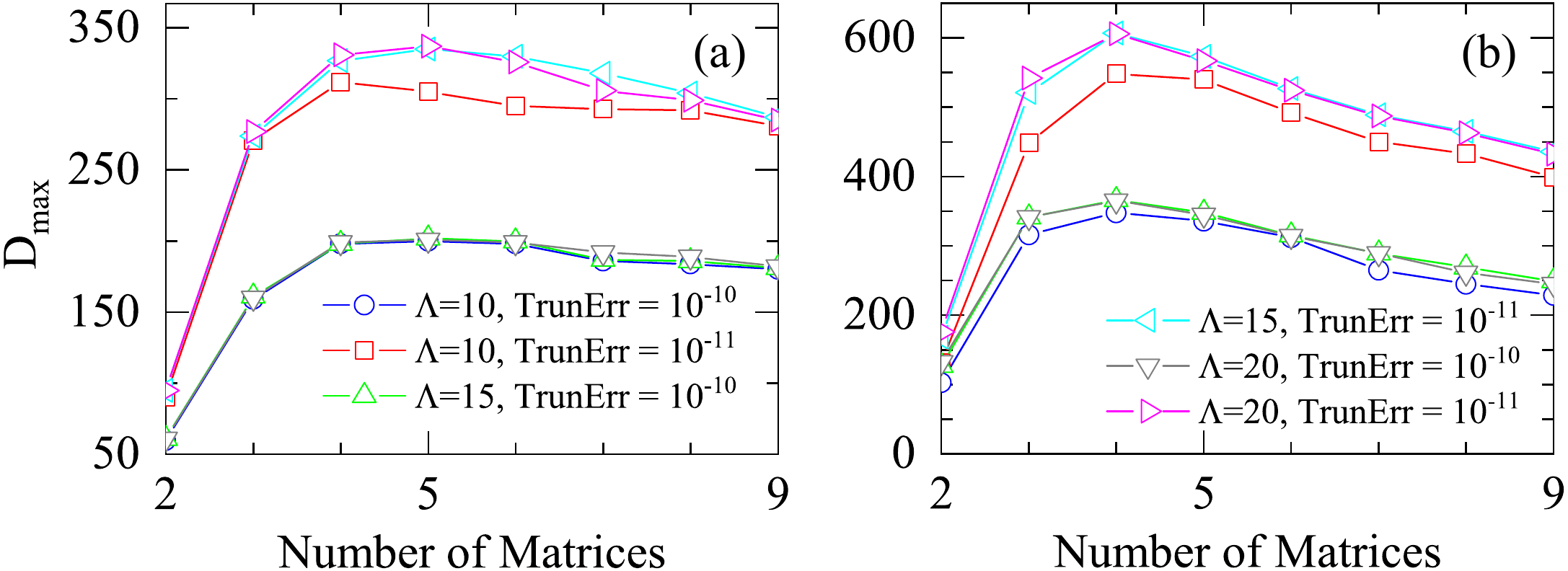}
    \caption{The maximal bond dimension $D_\mathrm{max}$ of MPS required to ensure the truncation error is less than $10^{-10}$ and $10^{-11}$ for the ground state when (a) $g=0.5$ and (b) $g=1.0$. }
    \label{fig:BMMSU2Bond}
\end{figure}

We first analyse the maximal bond dimension $D_\mathrm{{max}}$ of the MPS required in the final sweep procedure. $D_\mathrm{{max}}$ is determined by constraining the truncation error to remain below a chosen control threshold, referred to as TE. Here, we consider two values for TE, $10^{-10}$ and $10^{-11}$, to account for cases with varying truncation errors. Fig.~\ref{fig:BMMSU2Bond}(a) shows the $D_\mathrm{max}$ required for cases ranging from two to nine matrices, with the interaction strength $g$ fixed at $0.5$. Three representative values of $\Lambda = 10, 15, 20$ are chosen to address the $D_\mathrm{max}$ needed in different cutoff. As expected, the smaller TE needs larger $D_\mathrm{max}$ for all matrices. $D_\mathrm{max}$ saturates when the number of matrices exceeds four. Also, the $D_\mathrm{max}$ does not depend much on the cutoff $\Lambda$. 

To confirm the above observations in the strong coupling regime, we increase the coupling strength from $g = 0.5$ to $g = 1.0$. As a result, for fixed TE, $D_\mathrm{max}$ increases but the general dependence on the cutoff $\Lambda$ and number of matrices remains the same, in particular also for $g=1$, we see a saturation of $D_\mathrm{max}$ as the number of matrices increases, and it also remains nearly independent of $\Lambda$. The observed saturation with the number of matrices, along with the independence from the cutoff $\Lambda$, implies that the computational cost grows polynomially with $N$ and the number of matrices when using MPS to simulate SU$(2)$ matrix models in both moderate and strong coupling regimes.

\begin{figure}[htbp]
    \centering
    \includegraphics[width=0.9\linewidth]{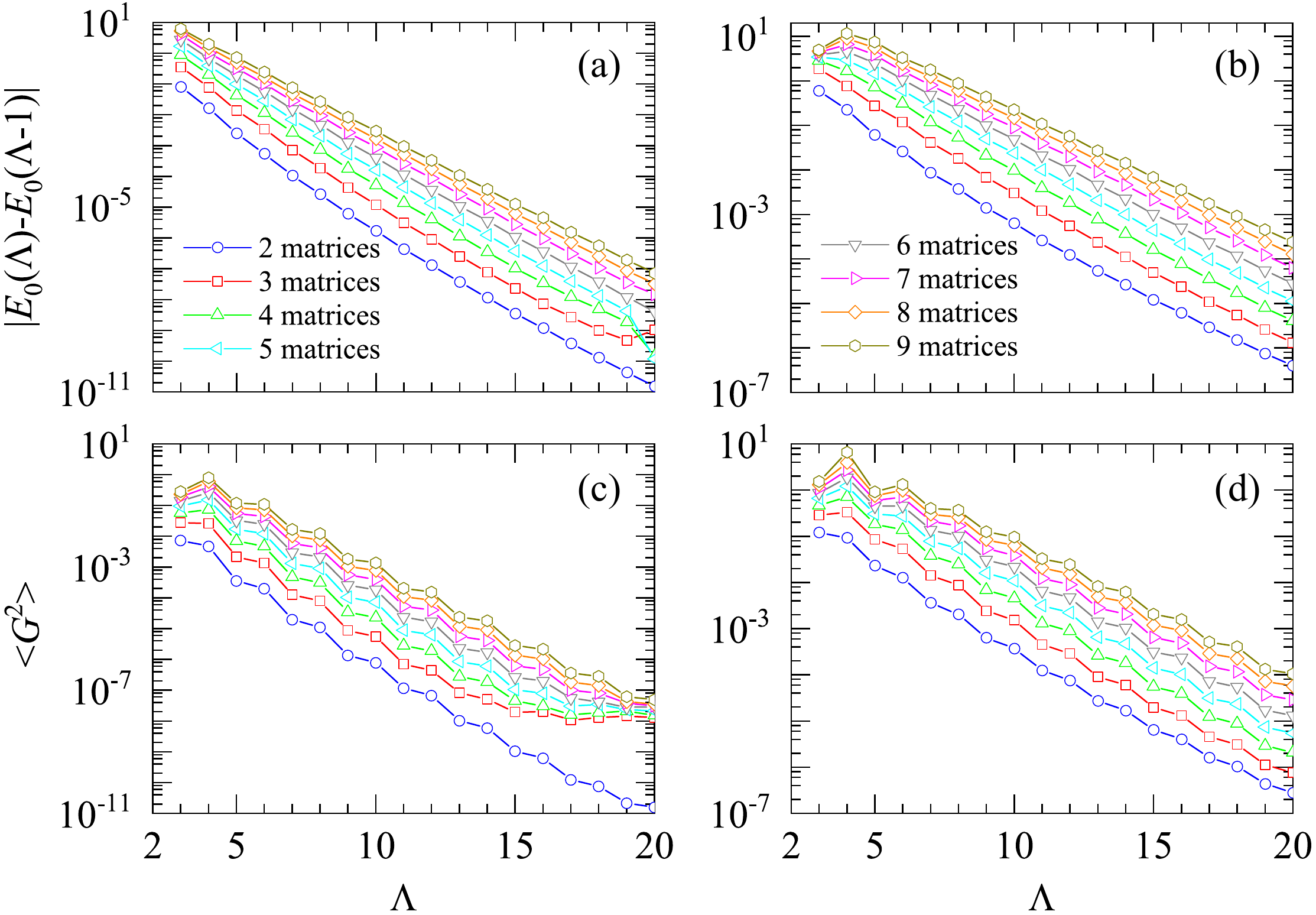}
    \caption{ The convergence of energy difference between the successive $\Lambda$ values, $\delta E(\Lambda) \equiv |E_0(\Lambda) - E_0(\Lambda-1)|$ and the gauge-invariance violation expectation value, $\langle G^2 \rangle$. The $\delta E(\Lambda)$ with $\Lambda$ for SU$(2)$ bosonic matrix models with (a) $g=0.5$ and (b) $g=1.0$. The $\langle G^2 \rangle$ with $\Lambda$ for (c) $g=0.5$ and (d) $g=1.0$. The truncation error in the simulation is fixed as $10^{-11}$. }
    \label{fig:GdE}
\end{figure}

We now examine the convergence of physical quantities, such as energy and the gauge-invariance violation expectation value $\langle G^2 \rangle$, with respect to the cutoff $\Lambda$. This convergence analysis serves two purposes. First, it allows us to evaluate and potentially mitigate the finite $\Lambda$ effects, where the behaviour of the energy as a function of $\Lambda$ provides a quantitative measure of these effects. Second, it is essential to monitor the variation in $\langle G^2 \rangle$, as sufficiently small values are crucial for ensuring meaningful results. Although no penalty term is applied, we expect $\langle G^2 \rangle$ is negligible for large enough $\Lambda$ which, however, should be numerically verified.

Figures~\ref{fig:GdE}(a-b) illustrates the variation of $\delta E(\Lambda) \equiv |E_0(\Lambda) - E_0(\Lambda-1)|$ (y-axis) with $\Lambda$ (x-axis) for the number of matrices ranging from two to nine. Fig.~\ref{fig:GdE}(a) shows the results for $g=0.5$, with the y-axis on a logarithmic scale. The nearly linear behaviour in the semi-logarithmic plot suggests that $\delta E(\Lambda)$ decays exponentially with $\Lambda$, demonstrating the fast convergence. Note that there is an abnormal convergence at $\delta E(\Lambda=20)$ for 3, 4, and 5 matrices, which arises due to the MPS stopping criterion being set as $10^{-8}$. When $g=1.0$, $\delta E(\Lambda)$ shown in Fig.~\ref{fig:GdE}(b) also decays exponentially with $\Lambda$, but with a smaller decay rate. 

\begin{figure}[htbp!]
    \centering
    \includegraphics[width=0.5\linewidth]{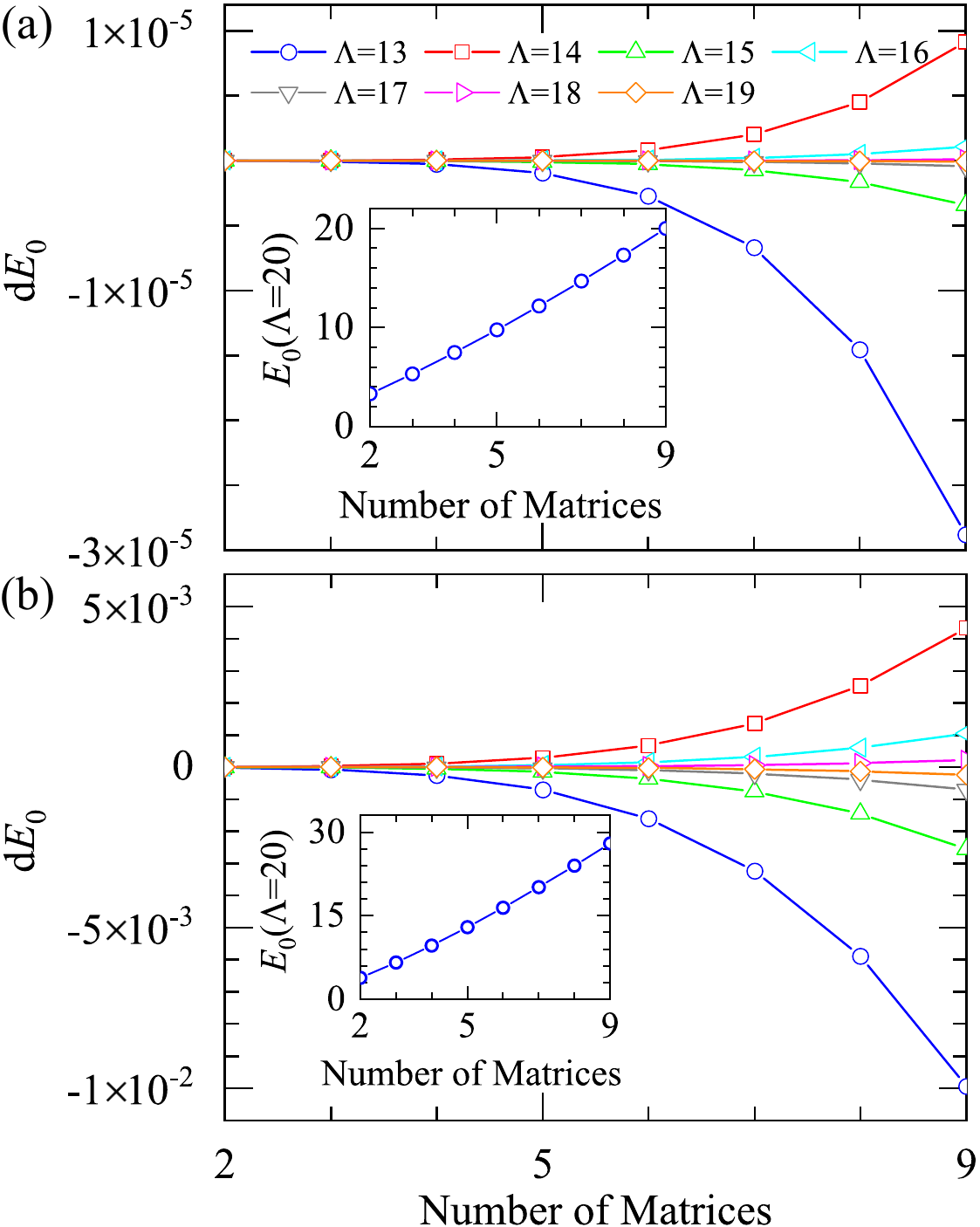}
    \caption{ The ground state energy difference $dE_{0}$ between the various $\Lambda$ and $\Lambda=20$ for SU$(2)$ bosonic matrices models with two to nine matrices. The $dE_{0}$ as a function of the number of matrices for (a) $g=0.5$ and (b) $g=1.0$. The insets in (a) and (b) show the ground state energy for $\Lambda=20$ from two to nine matrices at $g=0.5$ and $g=1.0$, respectively. }
    \label{fig:EdE}
\end{figure}

Figures~\ref{fig:GdE}(c-d) illustrates how $\langle G^2 \rangle$ (y-axis) varies with $\Lambda$ (x-axis), with Fig.~\ref{fig:GdE}(c) showing the interaction strength for $g=0.5$ and Fig.~\ref{fig:GdE}(d) for $g=1.0$. 
When the y-axis is displayed on a logarithmic scale, $\langle G^2 \rangle$ exhibits a linear decrease tendency with an even-odd oscillation pattern. 
This odd-even oscillation pattern stems from the effect of the hard cutoff of the bosonic excitations and vanishes when $\Lambda$ is large enough. 
A clear linear decrease of $\log(\langle G^2 \rangle)$ with $\Lambda$ is visible when separating the data into odd and even $\Lambda$, indicating a good exponential decaying behaviour of $\langle G^2 \rangle$ with $\Lambda$ for both $g=0.5$ and $g=1.0$.  
In Fig.~\ref{fig:GdE}(c), saturation occurs when $\langle G^2 \rangle$ reaches $\mathcal{O}(10^{-8})$, which is attributed to the MPS stopping criterion set at $10^{-8}$. 
The behaviour clearly suggest that the breaking of the singlet condition decays exponentially with $\Lambda$, confirming that finite cutoff effects can be neglected for the ground state for large $\Lambda$ without penalty terms.

In the final part of this subsection, we present the energy for different numbers of matrices. 
The energy differences, $dE_{0}$, between the ground states for various $\Lambda$ ($\Lambda = 13, 14, 15, 16, 17, 18, 19$) and $\Lambda = 20$ are shown in Fig.~\ref{fig:EdE}. 
As before, we consider both the moderate and strong coupling regimes, with Fig.~\ref{fig:EdE}(a) for $g=0.5$ and Fig.~\ref{fig:EdE}(b) for $g=1.0$. 
Comparing the two figures, we observe that the amplitude of $dE_{0}$ is greater for $g=1.0$ than for $g=0.5$. 
Additionally, $dE_{0}$ shows monotonic increase with the number of matrices for the same $\Lambda$ in both cases. 
For a fixed number of matrices, $dE_{0}$ approaches zero with oscillation between positive and negative values as $\Lambda$ increases. 
This shrinking oscillation is consistent with the behaviour we observed in Fig.~\ref{fig:GdE}(c-d). 
The insets in Fig.~\ref{fig:EdE}(a) shows the ground state energy for $\Lambda=20$ from two to nine matrices at $g=0.5$. 
As the number of matrices increases, the energy $E_{0}(\Lambda=20)$ grows almost linearly, a trend also found for $g=1.0$ shown in the inset of Fig.~\ref{fig:EdE}(b). 
Based on the observations in this subsection, we conclude that MPS can effectively simulate the SU$(2)$ bosonic matrix models with two to nine matrices, even at strong coupling.

\subsection{SU(N) bosonic matrix models}
We now want to evaluate how the current framework behaves under increasing $N$. We use the same strong coupling regime as in Ref.~\cite{Bodendorfer:2024egw} to benchmark the feasibility of using MPS to simulate matrix models. During the simulation, we found that the available bond dimension was insufficient to reach the fixed truncation error of $10^{-10}$ except for $N=2$. Therefore, we chose to use scaling analysis with the bond dimension $1/D$ to infer the energy and other physical observable in the $D \rightarrow \infty$ limit for $N=3, 4, 5$. The extrapolated bond dimension goes from 100 to 400 for $N = 3, 4$ and from 150 to 350 for $N=5$, both in steps of 50. For $N=6$,  we provided only a rough estimation for $D=200$ due to computational cost constraints.

\begin{figure}[htbp!]
    \centering
    \includegraphics[width=0.85\linewidth]{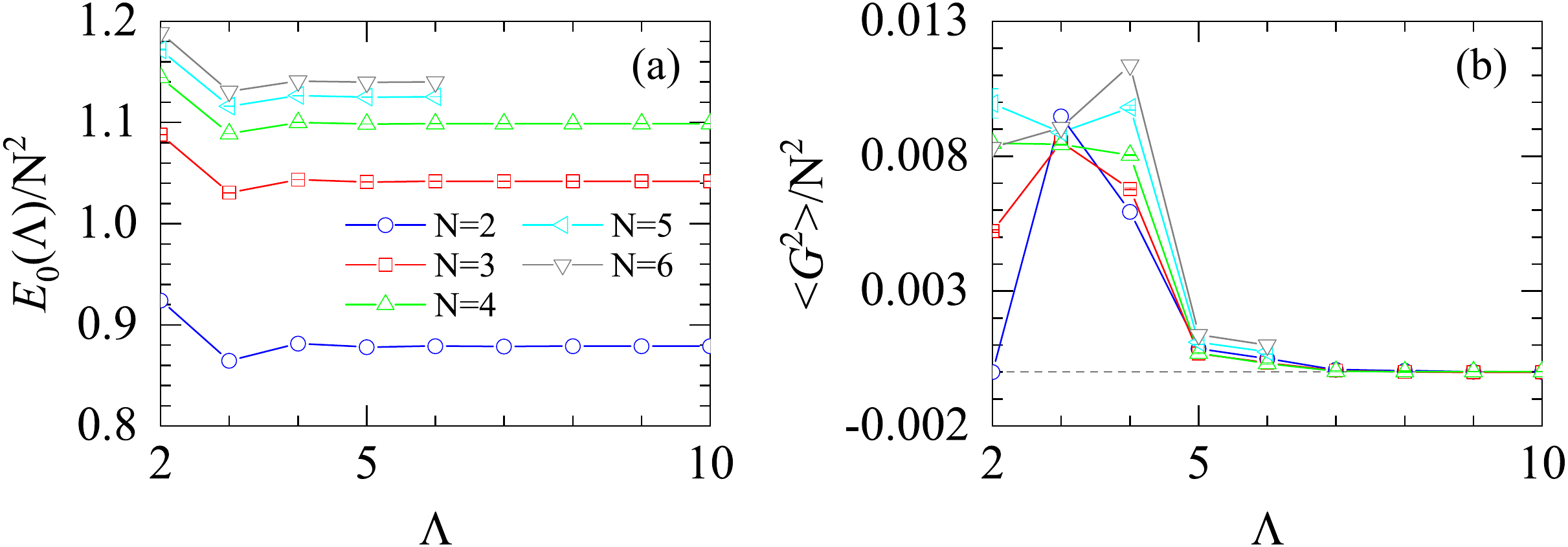}
    \caption{ Convergence of (a) the ground state energy $E_{0}(\Lambda)/N^{2}$ and (b) the gauge-invariance violation expectation value $\langle G^{2} \rangle / N^{2}$ as functions of $\Lambda$ for $N=2, 3, 4, 5,$ and $6$. Parameters are set with $m=1$ and the coupling $g$ such that $g^{2}N = 1$. For $N=2$, data are obtained with a truncation error kept below $10^{-10}$. For $N=3, 4,$ and $5$, data are extrapolated to the limit $D \rightarrow \infty$, with error bars recognized as the difference between two fitting functions. For $N=6$, data are presented with $D = 200$, constrained by computational cost. The dashed line in (b) indicates the zero level as a valid guide. }
    \label{fig:BMMSUNcutoffEG}
\end{figure}

Figure \ref{fig:BMMSUNcutoffEG} demonstrates the convergence of the ground state energy and the gauge-invariance violation expectation value for two matrices in the absence of penalty terms. To make direct comparison with the results reported in Ref.~\cite{Bodendorfer:2024egw}, we present $E_{0}(\Lambda)/N^{2}$ and $\langle G^{2} \rangle / N^{2}$ rather than $E_{0}(\Lambda)$ and $\langle G^{2} \rangle$ in this subsection. For $N=3$ and $N=4$, error bars are identified as the difference between quadratic and cubic fitting results, while for $N=5$, they are determined by the difference between linear and quadratic fits. As shown in Fig.~\ref{fig:BMMSUNcutoffEG}(a), $E_{0}(\Lambda)/N^{2}$ converges with even-odd oscillations as $\Lambda$ increases and eventually stabilizing, similar to the previous observations in SU(2) cases. 

To validate the ground state preserves gauge invariance, we further examine $\langle G^{2} \rangle / N^{2}$, shown in Fig.~\ref{fig:BMMSUNcutoffEG}(b). As indicated in Fig.~\ref{fig:BMMSUNcutoffEG}(b), $\langle G^{2} \rangle / N^{2}$ maintains a non-zero value at small $\Lambda$, gradually approaching zero as $\Lambda$ increases, suggesting full restoration of gauge invariance in the large-$\Lambda$ limit.

\begin{figure}[htbp!]
    \centering
    \includegraphics[width=0.55\linewidth]{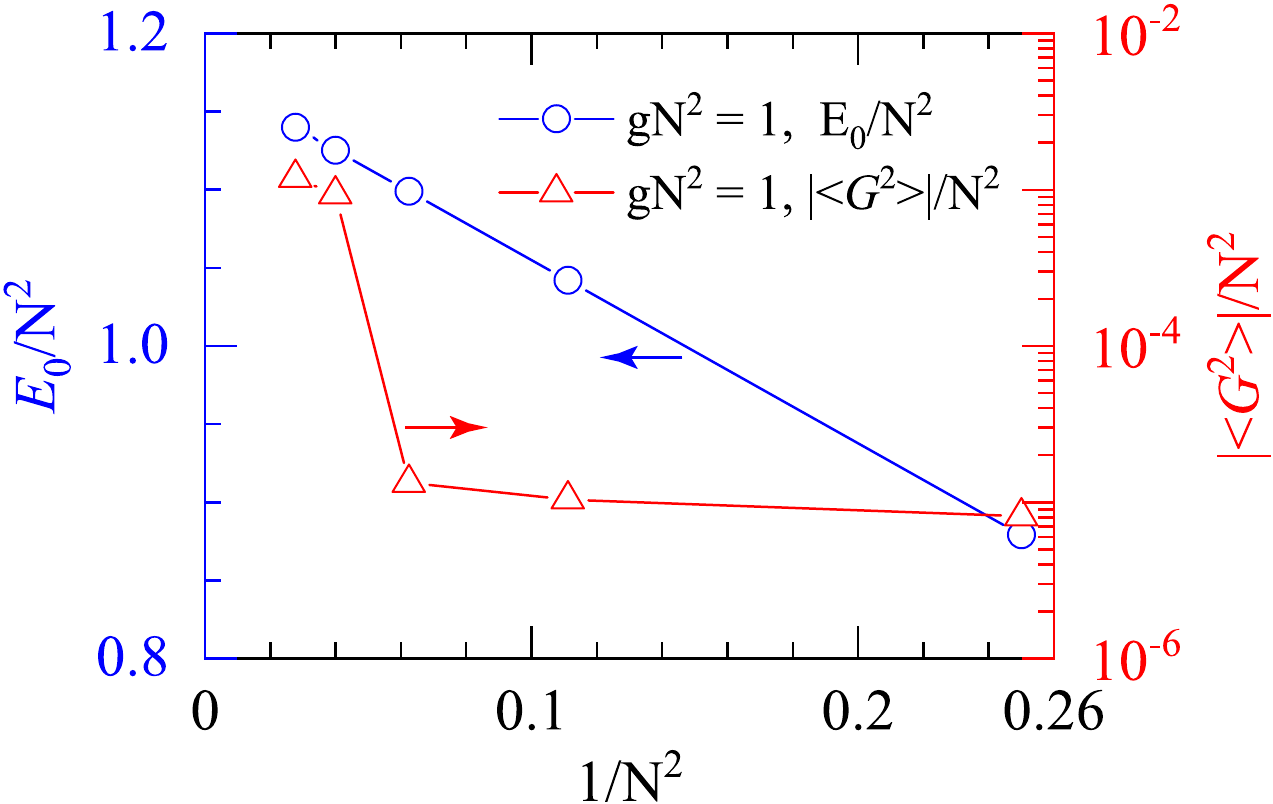}
    \caption{ Ground state energy, $E_{0}/N^{2}$ (left axis), and violation of gauge-invariant, $\langle G^{2} \rangle / N^{2}$ (right axis), shown as functions of $1/N^{2}$. Data points represent averages from the two largest $\Lambda$ values displayed in Fig.~\ref{fig:BMMSUNcutoffEG}. }
    \label{fig:BMMSUNEG}
\end{figure}

For two different SU(N) matrices, rather than extrapolating the data in Fig.~\ref{fig:BMMSUNcutoffEG} to infinite $\Lambda$, we use the average values from the last two $\Lambda$ in Fig.~\ref{fig:BMMSUNcutoffEG} to mitigate the effect of even-odd oscillations, providing an estimate for $D \rightarrow \infty$ and $\Lambda \rightarrow \infty$ limits. The ground state energy and deviation from gauge-invariance in these limits, denoted as $E_{0}/N^{2}$ and $\langle G^{2} \rangle / N^{2}$, are shown in Fig.~\ref{fig:BMMSUNEG} respectively.

Looking at Fig.~\ref{fig:BMMSUNEG}, we find that $\langle G^{2} \rangle / N^{2}$ grows slightly for $N = 5$ and $N=6$, mainly due to limiting $\Lambda$ to a maximum $6$ here. Nevertheless, $\langle G^{2} \rangle / N^{2}$ remains significantly smaller than the values reported in Ref.~\cite{Bodendorfer:2024egw}. Combined with the excellent agreement between the ground state energies we observe and those reported in Ref.~\cite{Bodendorfer:2024egw}, this suggests that the TNS framework we are using performs effectively across different values of $N$.

\subsection{SUSY Model}

We proceed by presenting results that include fermions. The supersymmetric SU$(N)$ matrix model that we mainly want to focus on was introduced in Sec.~\ref{sec:toyModel} with the Hamiltonian given in Eq.~\eqref{eq:HamiltonianBos}. In this subsection, we present the MPS simulation results for this model as illustrative example of SUSY matrix models.

\subsubsection{The model with N=2}

\begin{figure}[htbp!]
    \centering
    \includegraphics[width=0.5\linewidth]{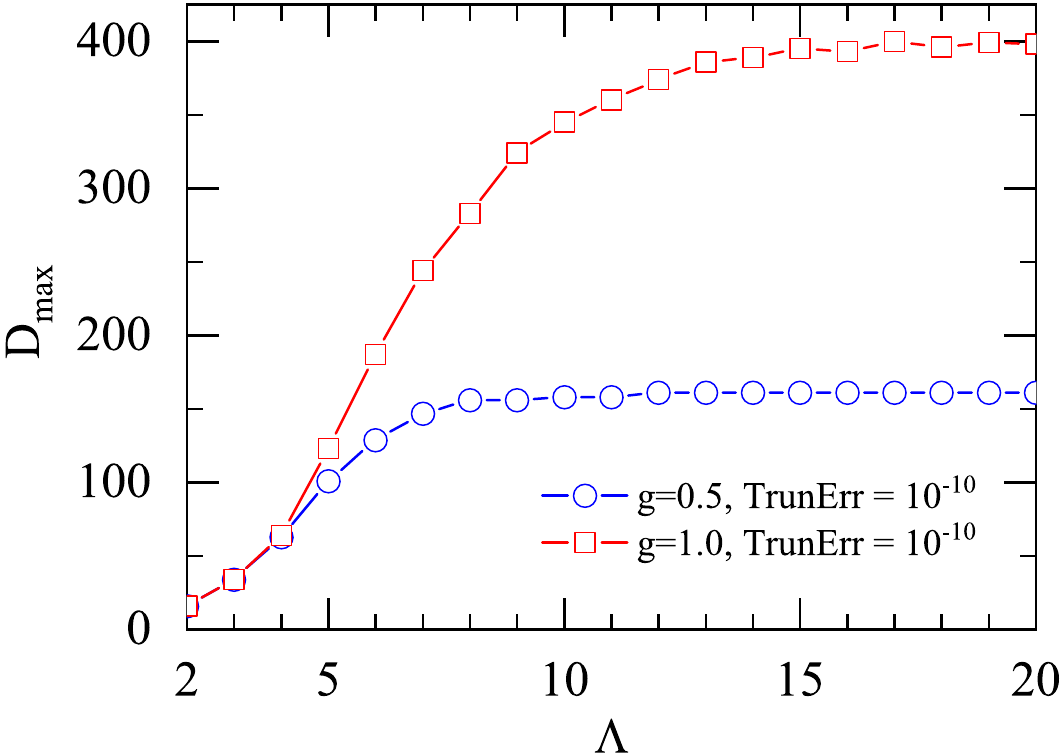}
    \caption{ Maximal bond dimension $D_\mathrm{max}$ required to maintain the truncation error below $10^{-10}$ for various $\Lambda$ in SU$(2)$ BMN models with two matrices. }
    \label{fig:BMNSU2Dmax}
\end{figure}

We begin by presenting results for SU$(2)$ with coupling constants $g=0.5$ and $g=1.0$. Using the same simulation strategy as for the SU$(2)$ bosonic matrix models, we fix the truncation error at $10^{-10}$ and show the maximum bond dimension $D_\mathrm{max}$ in Fig.~\ref{fig:BMNSU2Dmax}. We see that $D_\mathrm{max}$ increases with $\Lambda$ and eventually saturates for sufficiently large $\Lambda$. As before, we did not include the penalty terms to enforce the singlet constraint, as it is expected to be approximately satisfied with increasing $\Lambda$, as discussed in the previous section. We also omit any penalty terms that would fix angular momentum, expecting that the angular momentum expectation value will approach zero as $\Lambda$ increases. 

\begin{figure}[htbp!]
    \centering
    \includegraphics[width=0.5\linewidth]{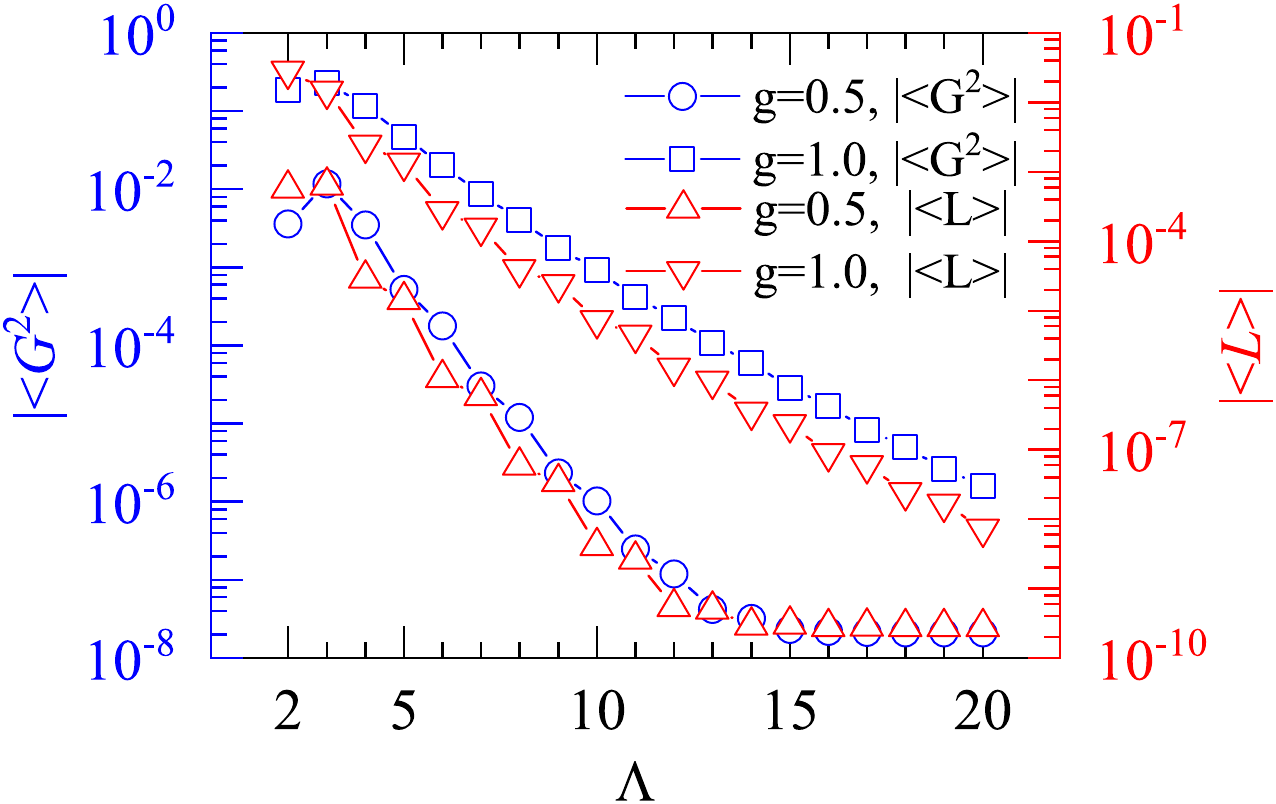}
    \caption{Violation of gauge invariance, $\langle G^2 \rangle$ (left y-axis) and absolute expectation value of angular momentum, $|\langle L \rangle|$ (right axis), for various $\Lambda$ in SU$(2)$ BMN models with two matrices.}
    \label{fig:BMNSU2G2L}
\end{figure}

Figure \ref{fig:BMNSU2G2L} shows the $\langle G^2 \rangle$ (left y-axis) for various $\Lambda$ with coupling constant $g = 0.5$ and $g = 1.0$. As $\Lambda$ increases, $\langle G^2 \rangle$ decays exponentially with odd-even oscillations, saturating when $\Lambda>13$ for $g = 0.5$ and continuing to decay for $g = 1.0$. In addition to $\langle G^2 \rangle$, we measured the absolute expectation value of angular momentum $|\langle L \rangle|$ (right axis). Similar to $\langle G^2 \rangle$, $|\langle L \rangle|$ decays exponentially with odd-even oscillations and saturates at $10^{-10}$, consistent with the truncation error set at $10^{-10}$. This observation of zero angular momentum observation aligns with previous expectations.

\begin{figure}[htbp!]
    \centering
    \includegraphics[width=0.55\linewidth]{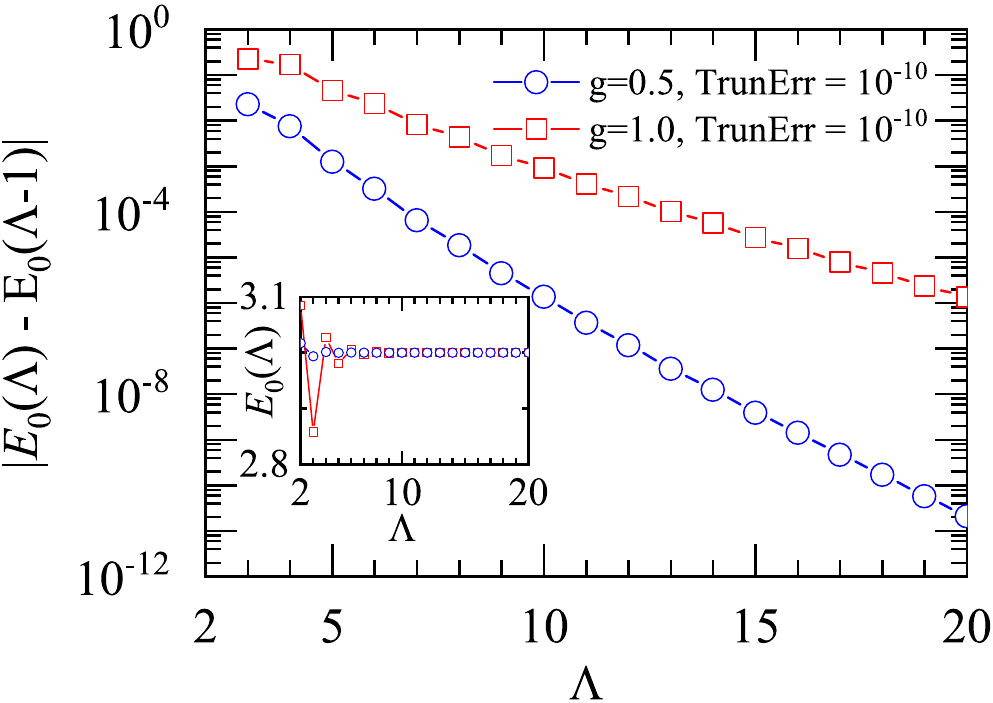}
    \caption{ Ground state energy difference, $\delta E_{\Lambda}\equiv|E_{0}(\Lambda)-E_{0}(\Lambda-1)|$, between successive $\Lambda$ values for SU$(2)$ BMN models with two matrices. The inset shows the ground state energy, $E_{\Lambda}$, for two matrices at $g=0.5$ and $g=1.0$, respectively. }
    \label{fig:BMNSU2EdE}
\end{figure}

Figure \ref{fig:BMNSU2EdE} reports the energy difference between the successive $\Lambda$ values, $\delta E_{0}(\Lambda) \equiv |E_{0}(\Lambda)-E_{0}(\Lambda-1)|$, as a function of $\Lambda$. As shown, $\delta E_{0}(\Lambda)$ decays exponentially and exhibits similar odd-even oscillatory behaviour as observed in bosonic matrix models. The inset shows the convergence of the ground state energy $E(\Lambda)$ with respect to $\Lambda$ and shows that the ground state energy converges faster for smaller coupling constants. Based on these results, we conclude that our TNS simulation framework produces very accurate results for the supersymmetric SU$(2)$ Yang-Mills matrix models with two matrices. The analysis of larger supersymmetric matrix models will be presented in future work. 

\subsection{Exploring low-energy physical observables}

In this subsection, we present and analyse some physical quantities of the models whose ground states we have computed in the previous sections. 
We focus on the toy model of 9 SU$(2)$ bosonic matrices with $m=1.0$ and $g=1.0$ as a representative example. 
We begin by introducing the boson number expectation value, followed by a discussion of entanglement properties, including the entanglement spectrum and entanglement entropy.

\subsubsection{Expectation values of bosons}

\begin{figure}[htbp!]
    \centering
    \includegraphics[width=0.55\linewidth]{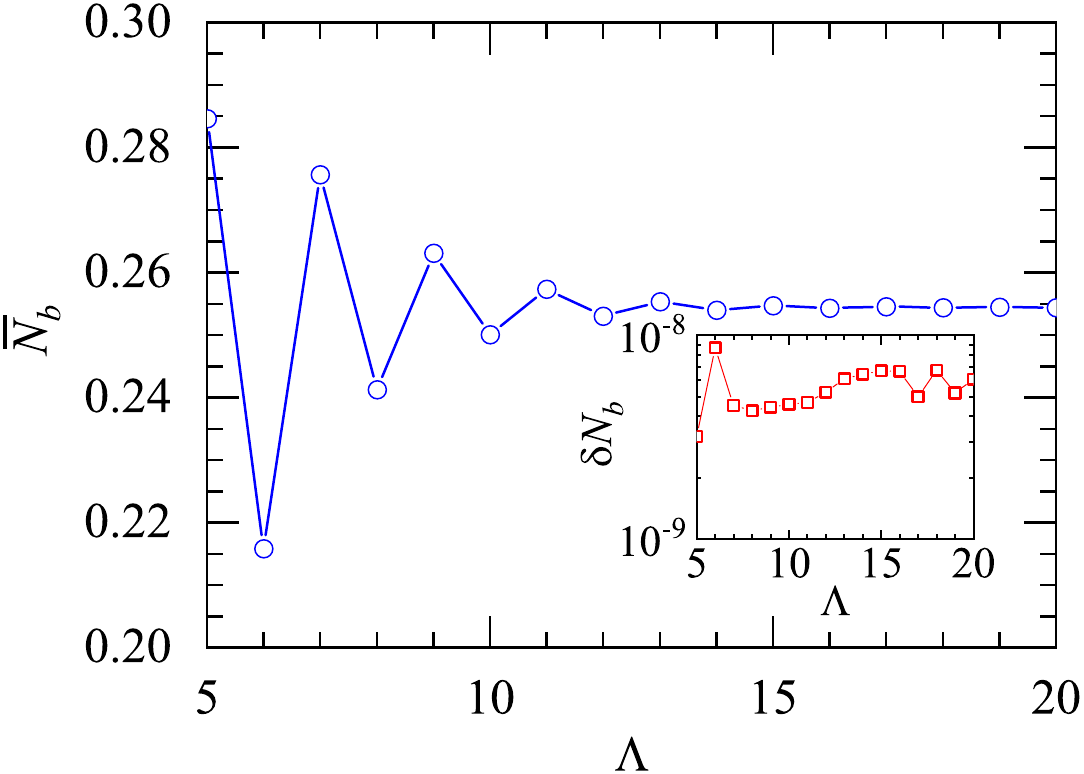}
    \caption{ Mean boson number expectation value per site, $\bar{N}_{b}$, as a function of $\Lambda$ for the nine SU$(2)$ bosonic matrix model with $m=1.0$ and $g=1.0$. The inset shows the maximum difference in expectation values across different sites as $\Lambda$ varies. }
    \label{fig:BMNSU2Nb}
\end{figure}

We begin by calculating the boson expectation values in the ground state, $\langle n_{i\alpha}\rangle_\Omega$,  
which are non-zero for any interaction constant $g>0$. Figure~\ref{fig:BMNSU2Nb} shows the mean vacuum boson number expectation value per site, $\bar{N}_{b} := \sum_{i,\alpha}\langle n_{i\alpha}\rangle_\Omega/27$, as a function of $\Lambda$. 
As $\Lambda$ increases, $\bar{N}_{b}$ exhibits an oscillatory behaviour with an odd-even pattern, where the oscillation amplitude decays approximately exponentially. 
The inset displays $\delta N_{b}$ for various $\Lambda$, where $\delta N_{b}$ represents the maximum variation in $N_{b}$ across all sites. 
For all cutoffs $\Lambda$, $\delta N_{b}$ remains between $10^{-9}$ and $10^{-8}$, indicating a uniform distribution of $N_{b}$ across sites. 
This uniform distribution is expected from the \ref{eq:HamiltonianBos}, which preserves permutation symmetry among the generators. 
Furthermore, we observed that $\bar{N}_{b}$ remains low for different $\Lambda$, indicating a large overlap between the ground state and the vacuum state, $\langle \Omega | \psi \rangle$. 
The low $\bar{N}_{b}$ also allows us to obtain accurate estimates of physical quantities with only a few $\Lambda$ values. 
It further matches the exponential decay observed in $\delta E(\Lambda)$ and $|\langle G^{2} \rangle|$, as the effect of the hard cutoff on occupation numbers is expected to decay exponentially with $\Lambda$ when $\Lambda>>\bar{N}_{b}$.

\subsubsection{Entanglement entropy and entanglement spectrum}

\begin{figure}[htbp!]
    \centering
    \includegraphics[width=0.5\linewidth]{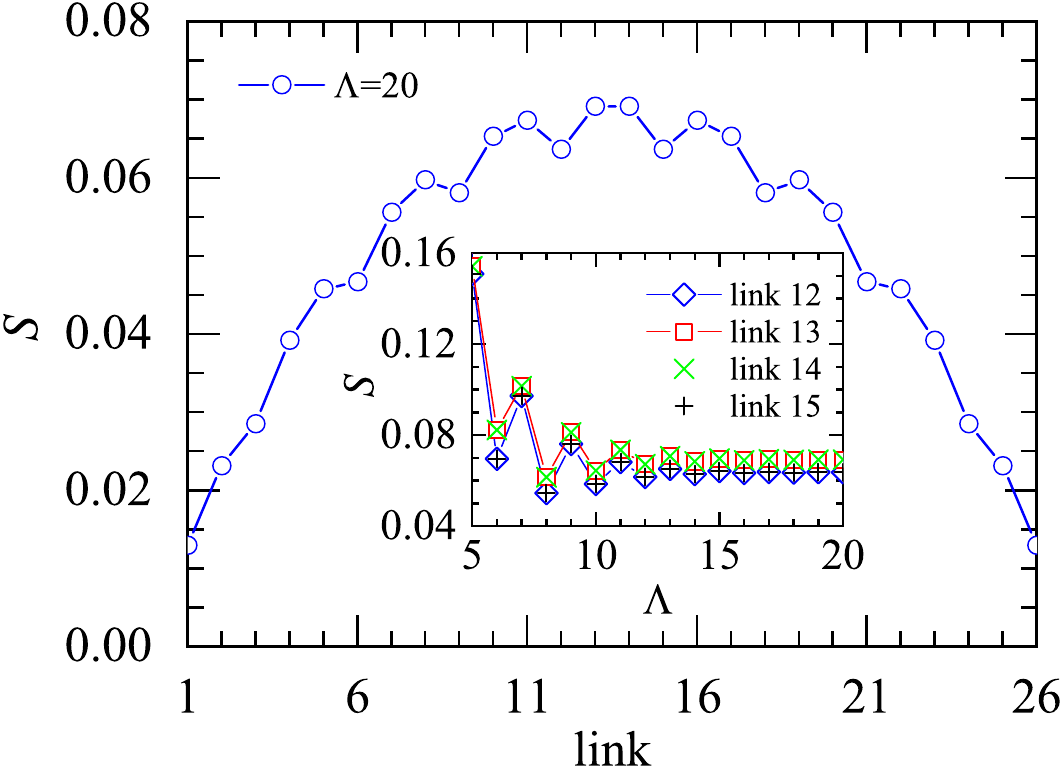}
    \caption{ Entanglement entropy $S$ of the ground state by cutting links for nine SU$(2)$ bosonic matrices with $m=1.0$ and $g=1.0$ at $\Lambda=20$. The inset shows the typical convergence behaviour of $S$ for several central links as a function of $\Lambda$.}
    \label{fig:BMMSU2Entropy}
\end{figure}

We proceed to present the entanglement properties of the ground state. 
Our TNS framework is particularly effective for extracting entanglement properties, including entanglement entropy and entanglement spectrum which were introduced in Sec.~\ref{sec:entanglement}. 
Figure \ref{fig:BMMSU2Entropy} shows the entanglement entropy $S$ of the ground state for all links connecting two sites and thereby the two regions on the two sites of the link. 
As shown in the inset, $S$ converges with $\Lambda$, and therefore we focus on $S$ at $\Lambda = 20$, where convergence has been achieved. 
We observe that $S$ displays a three-site oscillation pattern, consistent with the three bosonic degrees of freedom in each matrix. 
The entanglement entropy $S$ has a reflection symmetry about the central site, which is a consequence of the sequential layout scheme used in the mapping. 
We also observe that $S$ remains rather low, consistent with the low value of $\bar{N}_{b}$ and the  large overlap $\langle \Omega | \psi \rangle$, where $|\Omega \rangle$, the Fock vacuum is a product state and has no entanglement across the links. 

\begin{figure}[htbp!]
    \centering
    \includegraphics[width=0.5\linewidth]{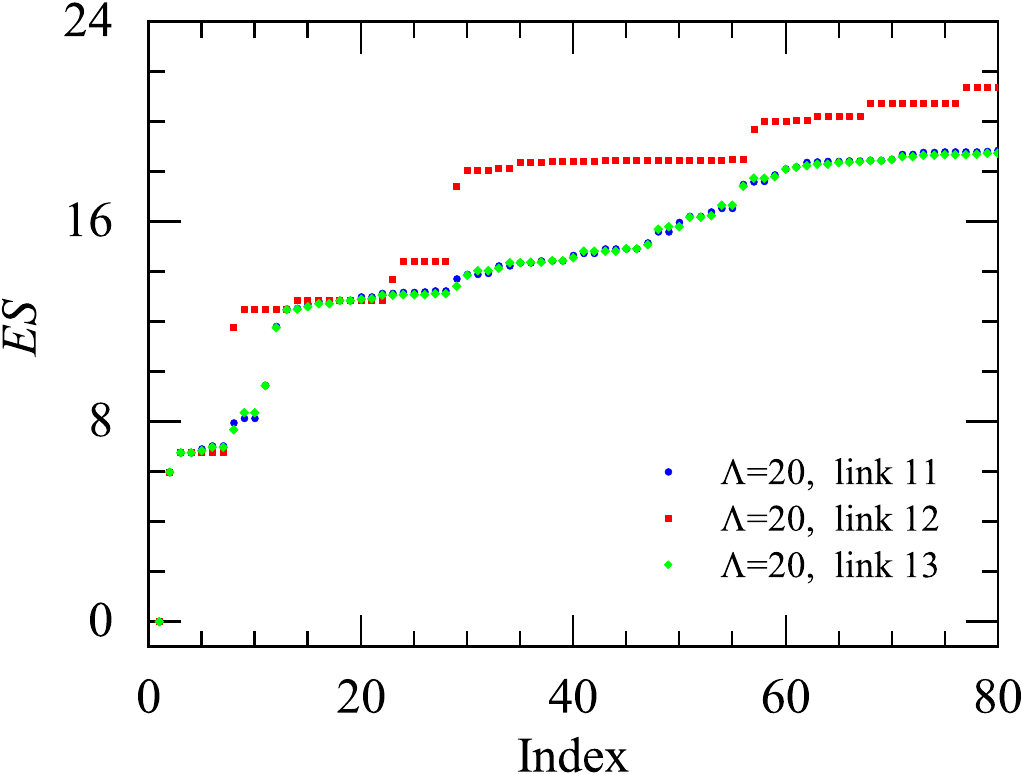}
    \caption{Entanglement spectrum, i.e. the collection of eigenvalues $\zeta_m$ given in Eq.~\eqref{eq:entEigenvalues}, of the ground state for typical links for nine SU$(2)$ bosonic matrices with parameters $m = 1.0$ and $g = 1.0$. The selected links at $\Lambda = 20$ illustrate representative results. }
    \label{fig:BMMSU2ES}
\end{figure}

Using the central links as examples, we present the entanglement spectrum of the ground state in Fig.~\ref{fig:BMMSU2ES}. Given that $S$ is symmetric around the central site, we select links 11 to 13 as representative examples. We set $\Lambda$ to 20, as $S$ has already converged at this value. A significant gap between the first two ES levels is observed, consistent with the previously identified behaviour of $S$ and $\bar{N}_{b}$. 

In summary, all above results highlight the efficiency of our TN simulation framework for computing entanglement properties.

We will elaborate on physical interpretation of these results in future work.  

\section{Conclusion and outlook}\label{sec:conclusion}

For the first time we successfully demonstrated the applicability of tensor network techniques, particularly matrix product states and the density matrix renormalization group, to matrix models. 
Our computations of ground states across varying numbers and sizes of matrices, as well as with increasing Fock space truncation, indicate that the computational cost scales polynomially with these quantities. 
This scaling behaviour, despite the complex non-local interactions across matrices and their elements, highlights the promise of tensor networks as a powerful tool for simulating matrix models and deepening our understanding of high-dimensional gauge theories, quantum gravity, and string theory.

Looking forward, we aim to refine our tensor network approach. 
One direction is to decompose the bosons into pseudo-bosons of smaller physical dimension, which would improve the reliability of MPS and DMRG methods. 
However, this approach introduces additional interaction terms and will make it necessary to find a balance between reducing local physical dimensions and the complexity of the interaction.
The possibility of using local basis optimization (LBO) technique~\cite{zhang1998density,iskin2005bcs} and single-site DMRG methods~\cite{hubig2015strictly,gleis2023controlled} can also be explored to improve the efficiency. 
Another promising avenue involves numerically imposing the SU$(N)$ gauge constraint within the free theory.
Although it demands significant preliminary effort, this approach can greatly reduce the state space at a fixed energy, enabling exploration at higher energy levels.

Beyond numerical and computational advancements, we will explore the physical implications of these models. 
This includes constructing excited states, mapping out phase diagrams, investigating potential phase transitions, and observing the thermodynamics of matrix models —- particularly as they relate to gravity and black holes. 
Additionally, we plan to investigate the feasibility of 2D tensor network structures, which may better capture the physics of matrix models. 
We also intend to explore synergies with quantum computing, especially for regimes where tensor network techniques face limitations.

\vspace{3cm}

\section*{Acknowledgments}

Y.~G.~is grateful to Dr. Zongsheng Zhou for the helpful discussions.
E.~R.~acknowledges Sheng-Hsuan Lin and Guillermo Preisser for important feedback on the manuscript. We also thank Stefan Kühn for helpful discussions and carefully checking the manuscript.  
This work is funded by the European Union's Horizon Europe Framework Programme
(HORIZON) under the ERA Chair scheme with grant agreement {no.~101087126}. 
This work is supported with funds from the Ministry of Science, Research and
Culture of the State of Brandenburg within the Centre for Quantum Technology and
Applications (CQTA).

\begin{center}
    \includegraphics[width = 0.16\textwidth]{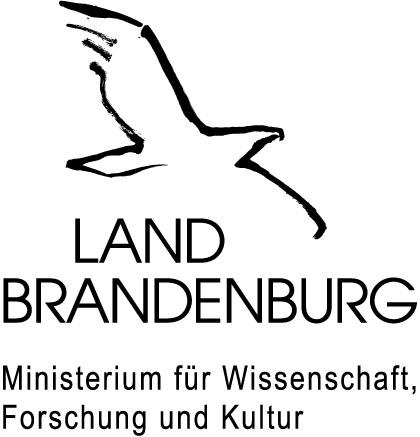}
\end{center}

\newpage

\appendix

\section{Expanding the fields in su(N) generators}\label{app:su(N)expansion}

We can expand the hermitian operators $X_i$ and $\xi_i$ in $\mathfrak{su}(N)$ generators
\begin{align} \label{eq:SUNexpansion}
    X_i = \sum_{\alpha=1}^{N^2 - 1} X_{i\alpha} \tau_\alpha\,, \qquad P_i = \sum_{\alpha=1}^{N^2 - 1} P_{i\alpha} \tau_\alpha\,, \qquad \xi = \sum_{\alpha=1}^{N^2-1} \xi_\alpha \tau_\alpha\,,
\end{align}
where we choose the normalisation $\Tr(\tau_\alpha\tau_\beta) = \delta_{\alpha\beta}$ and $[\tau_\alpha,\tau_\beta] = i \sum_\gamma f_{\alpha\beta\gamma} \tau_\gamma$ as our convention for the $\mathfrak{su}(N)$ structure constants $f_{\alpha\beta\gamma}$, which are given for arbitrary $N$ in the following App.~\ref{app:structureConst}. The expansions coefficients fulfil the usual canonical relations 
\begin{align}
    \{\xi_\alpha^\dagger,\xi_\beta\} = \delta_{\alpha\beta}\,,\quad [X_{i\alpha},P_{j\beta}] = i \delta_{ij}\delta_{\alpha\beta}\,.
\end{align}
For each bosonic degree of freedom $(i,\alpha)$ we can define the lowering and raising operators
\begin{align}
    a_{i\alpha} &= \sqrt{\frac{m}{2}} X_{i\alpha} + \frac{i}{\sqrt{2m}} P_{i\alpha}\,,\qquad
    a_{i\alpha}^\dagger = \sqrt{\frac{m}{2}} X_{i\alpha} - \frac{i}{\sqrt{2m}} P_{i\alpha}\,,
\end{align}
with commutation relations
\begin{equation}
    \left[a_{i\alpha},a_{j\beta}^\dagger\right] = \delta_{ij}\delta_{\alpha\beta}\,. 
\end{equation}
and the number operator 
\begin{equation}
n_{i\alpha} = a_{i\alpha}^\dagger a_{i\alpha}\,.
\end{equation}
For completeness, we also give the inverse relation 
\begin{align}
    X_{i\alpha} &= \frac{1}{\sqrt{2m}} \left(a_{i\alpha} + a_{i\alpha}^\dagger\right)\,,\qquad P_{i\alpha} = i\sqrt{\frac{m}{2}} \left(a_{i\alpha}^\dagger - a_{i\alpha}\right)\,.
\end{align}

\section{The su(N) Lie algebra basis and structure constants}\label{app:structureConst}

The explicit form of the structure constants of $\mathfrak{su}(N)$ -- the Lie algebra of the SU$(N)$ group -- are base dependent. We want to choose a basis of $N\times N$ hermitian matrices $\{\tau_\alpha,\alpha=1,\dots,N^2-1\}$ that is orthonormal with respect to the trace inner product $\langle A,B\rangle = \Tr A^\dagger B$, i.e. $\Tr(\tau_\alpha\tau_\beta) = \delta_{\alpha\beta}$. Such a basis is given by \cite{Bossion:2021zjn}
\begin{itemize}
    \item $N(N-1)/2$ symmetric matrices 
    \begin{equation}
        \tau_{\alpha_{mn}} = \frac{1}{\sqrt{2}} \left(\ket{m}\bra{n} + \ket{n}\bra{m}\right)\,
    \end{equation}
    with $N\ge m > n \ge 1$,
    
    \item $N(N-1)/2$ antisymmetric matrices
    \begin{equation}
        \tau_{\beta_{mn}} = \frac{i}{\sqrt{2}} \left(\ket{m}\bra{n} - \ket{n}\bra{m}\right)\
    \end{equation}
    again with $N\ge m > n \ge 1$, and
    
    \item $N-1$ diagonal matrices 
    \begin{equation}
        \tau_{\gamma_{n}} = \frac{1}{\sqrt{n(n-1)}}\left( \sum_{l=1}^{n-1} \ket{l}\bra{l} +(1-n) \ket{n}\bra{n}\right)\,
    \end{equation}
    for $N\ge n \ge 2$.  
\end{itemize}
We can map the three above indices to a single index $\alpha$ via 
\begin{equation}
\begin{split}
    \alpha_{mn} &\mapsto m^2 + 2(n-m) - 1\,,\\
    \beta_{mn} &\mapsto  m^2 + 2(n-m)\,,\\
    \gamma_n &\mapsto n^2-1\,.
\end{split}
\end{equation}

\noindent
In case of $N=2$ this construction gives the normalised Pauli matrices
\begin{align}
    \tau_1 &= \frac{1}{\sqrt{2}} \,\sigma_x = \frac{1}{\sqrt{2}} \begin{pmatrix}
        0 & 1 \\
        1 & 0
    \end{pmatrix}\,,\\
    \tau_2 &= \frac{1}{\sqrt{2}}\, \sigma_y = \frac{1}{\sqrt{2}} \begin{pmatrix}
        0 & -i \\
        i & 0
    \end{pmatrix}\,,\\
    \tau_3 &= \frac{1}{\sqrt{2}} \,\sigma_z = \frac{1}{\sqrt{2}} \begin{pmatrix}
        1 & 0 \\
        0 & -1
    \end{pmatrix}\,,    
\end{align}
and for $N=3$ one obtains $\tau_\alpha = \frac{\lambda_\alpha}{\sqrt{2}}$, where $\lambda_\alpha$ are the so-called Gell-Mann matrices \cite{PhysRev.125.1067}. 

If we define the structure constants as 
\begin{equation}
    [\tau_\alpha,\tau_\beta] = i \sum_\gamma f_{\alpha\beta\gamma} \tau_\gamma
\end{equation}
then the non-zero structure constants are given by \cite{Bossion:2021zjn} 
\begin{align}
    f_{\alpha_{mn}\alpha_{km} \beta_{kn}} &= f_{\alpha_{mn}\alpha_{mk} \beta_{kn}} = f_{\alpha_{mn}\alpha_{kn} \beta_{km}} = \frac{1}{\sqrt{2}}\,,\\
    f_{\beta_{mn}\beta_{kn} \beta_{km}} &= \frac{1}{\sqrt{2}}\,,\\
    f_{\beta_{mn}\beta_{mn} \gamma_{n~}} &= -\sqrt{\frac{n-1}{n}}\,,\quad f_{\beta_{mn}\beta_{mn} \gamma_{m}} = \sqrt{\frac{m}{m-1}}\\
    f_{\beta_{mn}\beta_{mn} \gamma_{k~}} &= \sqrt{\frac{1}{k(k-1)}}\,, \quad \text{for} \quad n<k<m\,.
\end{align}

\section{State space of the minimal BMN-like model}\label{app:statespace}

The space of bosonic excitations is created from the space of $2(N^2-1)$ free quantum harmonic oscillators which is also called Fock space. As a reminder, the Fock vacuum 
\begin{equation}
    \ket{\Omega_\mathrm{free}}\,
\end{equation}
is defined as the unique state that is annihilated by all lowering operators $a_{i\alpha}$,
\begin{equation}
    a_{i\alpha} \ket{\Omega_\mathrm{free}}=0\,.
\end{equation}
Fock states build the basis of the Fock space and are created by consecutive application of creation operators $a_{i\alpha}^\dagger$ onto the Fock vacuum
\begin{equation}
    \ket{\{n_{i\alpha}\}} \equiv \otimes_{i\alpha} \ket{n_{i\alpha}}_{i\alpha} = \left(\prod_{i\alpha}\frac{\left(a^{\dagger}_{i\alpha}\right)^{n_{i\alpha}}}{\sqrt{n_{i\alpha}!}}\right) \ket{\Omega_{\mathrm{free}}}\,.
\end{equation}
They are eigenstates of the number operators $n_{i\alpha} = a_{i\alpha}^\dagger a_{i\alpha}$ and, thus, eigenstates of the non-interacting bosonic Hamiltonian, $H_X|_{g=0}$\,. The eigenstates of the interacting theory can be constructed as linear combinations of Fock states. 

 On the Fock space the ladder operators can be written as 
\begin{equation}
  a_{i\alpha} = \sum_{m=0}^\infty \sqrt{m+1} \ket{m}_{i\alpha} \bra{m+1}_{i\alpha}\quad,\qquad a_{i\alpha}^\dagger = \sum_{m=0}^\infty \sqrt{m+1} \ket{m+1}_{i\alpha} \bra{m}_{i\alpha}\,
\end{equation}
and the number operator is
\begin{equation}
    n_{i\alpha} = \sum_{n=0}^\infty n \,\ket{n}_{i\alpha}\bra{n}_{i\alpha}\,.
\end{equation}
The Fock space is infinite dimensional whereas in a simulation we can only handle finitely dimensional Hilbert spaces. We can tackle the problem by truncating the Fock space and restrict $n_{i\alpha}$ for every $(i,\alpha)$ such that we only allow $\Lambda$ states: 
\begin{equation}
    0\le n_{i\alpha} < \Lambda-1\,. \label{eq:truncation}
\end{equation}
The ladder operators and the number operator are then
\begin{align}
    a^{\dagger}_{i,\alpha} &= \sum_{n=0}^{\Lambda-2} \sqrt{n+1} \ket{n+1}_{i,\alpha} \bra{n}_{i,\alpha} \,,\\
    a_{i,\alpha} &= \sum_{n=0}^{\Lambda-2} \sqrt{n+1} \ket{n}_{i,\alpha} \bra{n+1}_{i,\alpha}\,,\\
    \hat{n}_{i,\alpha} &= \sum_{n=1}^{\Lambda-1} n\, \ket{n}_{i,\alpha} \bra{ n}_{i,\alpha}\,.
\end{align}
The fermionic excitations are much simpler. We can again consider that there is a vacuum of the non-interacting theory which is annihilated by the fermionic modes $\xi_\alpha$, i.e. 
\begin{equation}
    \forall_\alpha\ \xi_\alpha \ket{\Omega_{\mathrm{free}}} = 0\,,
\end{equation}
where $\ket{\Omega_{\mathrm{free}}}$ can be regarded as the same unique vacuum as for the bosons. Excitations are now obtained by acting with $\xi^\dagger_\alpha$'s. 
However, since $(\xi^\dagger_\alpha)^2 = 0$ each individual fermion excitation can appear at most once. For $N^2-1$ different fermions there are $2^{N^2-1}$ states. For example, for $N=2$, there are three fermions $\xi_1,\xi_2,\xi_3$ and eight states
\begin{align}\nonumber
    \ket{\Omega_{\mathrm{free}}}\,&,&\quad \xi_1^\dagger\ket{\Omega_{\mathrm{free}}}\,&,&\quad \xi_2^\dagger\ket{\Omega_{\mathrm{free}}}\,&,&\quad \xi_3^\dagger\ket{\Omega_{\mathrm{free}}}\,&,\\
    \xi_1^\dagger\xi_2^\dagger\ket{\Omega_{\mathrm{free}}}\,&,&\quad\xi_1^\dagger\xi_3^\dagger\ket{\Omega_{\mathrm{free}}}\,&,&\quad\xi_2^\dagger\xi_3^\dagger\ket{\Omega_{\mathrm{free}}}\,&,&\quad\xi_1^\dagger\xi_2^\dagger\xi_3^\dagger\ket{\Omega_{\mathrm{free}}}\,&.
\end{align}
The fermions only play a secondary role for the effort to numerically simulate matrix models in the Hamiltonian picture. Each fermion only contributes a two dimensional state space such that there is in particular no need for a truncation. 

In the explicit realisation of fermions, it is important to mind the anti-commutation relation between them. E.g. following the Jordan-Wigner transformation, when mapping $K$ fermions to the state space of $K$ qubits (two level systems), then a possible representation of the fermionic generator is 
\begin{equation}
    \xi_i^\dagger = \begin{pmatrix}
        1 & 0 \\
        0 & -1
    \end{pmatrix}_1 \otimes \dots \otimes \begin{pmatrix}
        1 & 0 \\
        0 & -1
    \end{pmatrix}_{i-1} \otimes \begin{pmatrix}
        0 & 0 \\
        1 & 0
    \end{pmatrix}_i \otimes \begin{pmatrix}
        1 & 0 \\
        0 & 1
    \end{pmatrix}_{i+1} \otimes \dots \otimes \begin{pmatrix}
        1 & 0 \\
        0 & 1
    \end{pmatrix}_{K}\,.
\end{equation}

\section{Matrix product states and operators}\label{app:MPSandMPO}

For systems with $L$ $d$-dimensional sites, a quantum state can be written as
\begin{equation}
    | \psi \rangle = \sum_{\{\sigma_{i}, \alpha_{i}\}} A^{1, \sigma_{1}}_{\alpha_{0} \alpha_{1}} A^{2, \sigma_{2}}_{\alpha_{1} \alpha_{2}} \cdots A^{L-1, \sigma_{L-1}}_{\alpha_{L-2} \alpha_{L-1}} A^{L, \sigma_{L}}_{\alpha_{L-1} \alpha_{0}}|\sigma_{1}\sigma_{2}\cdots \sigma_{L-1} \sigma_{L} \rangle,
    \label{eq:MPS}
\end{equation}
where $|\sigma_{1}\sigma_{2} \cdots \sigma_{L-1}\sigma_{L} \rangle$ are product states of local basis vectors $| \sigma_{i} \rangle$. The collection of tensors $A$ is called matrix product state~\cite{ostlund1995thermodynamic,xiang2023density}. 
For an MPS with open (periodic) boundary conditions, the index $\alpha_{0}$ is set to $1$ (greater than $1$). 
$A^{i,\sigma_{i}}_{\alpha_{i-1},\alpha_{i}}$ is a parametrized rank-3 tensor located at the $i$th site. 
The index $\sigma_{i}$ runs from $1$ to $d$, which for example for our bosonic degrees of freedom will be given by the truncation $\Lambda$ of the local Fock spaces introduced before, and $2$ for the spin-1/2 systems. The index $\alpha_{i}$ ($\alpha_{i-1}$) range from $1$ to $D_{i}$ ($D_{i-1}$), where $D_{i}$ ($D_{i-1}$) is referred as the \textit{bond dimension} and gives the dimension of virtual spaces between the physical sites. The bond dimension, usually set to a site-independent value $D$, serves as a key adjustable parameter to control the number of parameters and tune the performance of the MPS ans\"atze.

An MPS can be represented with a diagrammatic notation. After writing the state coefficient as
\begin{equation}
    \psi^{\sigma_{1}\cdots\sigma_{L}} \equiv \sum_{\{\alpha_{i}\}} A^{1, \sigma_{1}}_{\alpha_{0} \alpha_{1}} A^{2, \sigma_{2}}_{\alpha_{1} \alpha_{2}} \cdots A^{L-1, \sigma_{L-1}}_{\alpha_{L-2} \alpha_{L-1}} A^{L, \sigma_{L}}_{\alpha_{L-1} \alpha_{0}},
\end{equation}
the diagrammatic notation is
\begin{equation}\label{eq:MPSgraph}
\centering
  \begin{tikzpicture}[every node/.style={scale=1},scale=0.5]
    \draw (-13.25, 0) node {$\psi^{\sigma_{1}\cdots\sigma_{L}}$} (-11.5, 0) node {$=$};
    \draw (-10,0)--(-11,0) (5,0)--(6,0);
    \draw (-9.5,0.5)--(-9.5,1.25) (-9.5,0)circle(0.5) (-9,0)--(-7.5,0);
    \draw (-9.5,1.25) node[above] {$\sigma_{1}$} (-9.5,-0.5)node[below]{$A^{1}$};
    \draw (-7,0.5)--(-7,1.25) (-7,0)circle(0.5) (-6.5,0)--(-5,0);
    \draw (-7,1.25) node[above] {$\sigma_{2}$} (-7,-0.5)node[below]{$A^{2}$};
    \draw (-4.5,0.5)--(-4.5,1.25) (-4.5,0)circle(0.5) (-4,0)--(-3,0);
    \draw (-4.5,1.25) node[above] {$\sigma_{3}$} (-4.5,-0.5)node[below]{$A^{3}$};
    \draw (-2.15,0) node {$\cdots$} (-1.5,0)--(-0.5,0);
    \draw (0,1.25) node[above] {$\sigma_{i}$} (0,-0.5)node[below]{$A^{i}$};
    \draw (0,0.5)--(0,1.25) (0,0) circle(0.5) (0.5,0)--(1.5,0);
    \draw (2.35,0) node {$\cdots$} (3,0)--(4,0);
    \draw (4.5,0.5)--(4.5,1.25) node[above] {$\sigma_{L}$} (4.5,0) circle(0.5) (4.5,-0.5)node[below]{$A^{L}$} (6.5, 0);
  \end{tikzpicture}
\end{equation}
for an MPS with open boundary conditions.
Here, the empty circle attached with three indices represent the individual rank-3 tensor. For $i$th and $i+1$th local tensors, the edge connecting them means the summation over the index $\alpha_{i}$, while the dangling open legs upwards indicate the local physical degrees of freedom. The dimension of the open legs on the boundary is set to $1$, maintained here for convenience in the subsequent discussion.

Naturally, operators can be written with the matrix product form similar to MPS. This form is called matrix product operators (MPO)~\cite{pirvu2010matrix,chan2016matrix,xiang2023density}. Consider a Hamiltonian $H$ as an example which can be written as
\begin{equation}
    H = \sum_{\{\sigma_{i}, \sigma^{\prime}_{i}, \alpha_{i}\}} W^{1, \sigma^{\prime}_{1} \sigma_{1}}_{\alpha_{0} \alpha_{1}} W^{2, \sigma^{\prime}_{2} \sigma_{2}}_{\alpha_{1} \alpha_{2}} \cdots W^{L, \sigma^{\prime}_{L} \sigma_{L}}_{\alpha_{L-1} \alpha_{0}} |\sigma^{\prime}_{1}\sigma^{\prime}_{2}\cdots \sigma^{\prime}_{L} \rangle \langle \sigma_{1}\sigma_{2}\cdots \sigma_{L}|,
    \label{eq:MPO}
\end{equation}
where the coefficient $H^{\sigma^{\prime}_{1} \cdots \sigma^{\prime}_{L}}_{\sigma_{1} \cdots \sigma_{L}}$ in front of the basis elements $|\sigma^{\prime}_{1}\sigma^{\prime}_{2}\cdots \sigma^{\prime}_{L} \rangle \langle \sigma_{1}\sigma_{2}\cdots \sigma_{L}|$ is the matrix product of the series of rank-4 tensors $ W^{i, \sigma^{\prime}_{i} \sigma_{i}}_{\alpha_{i-1} \alpha_{i}}$, that is
\begin{equation}
    H^{\sigma^{\prime}_{1} \cdots \sigma^{\prime}_{L}}_{\sigma_{1} \cdots \sigma_{L}} \equiv \sum_{\{\sigma_{i}, \sigma^{\prime}_{i}, \alpha_{i}\}} W^{1, \sigma^{\prime}_{1} \sigma_{1}}_{\alpha_{0} \alpha_{1}} W^{2, \sigma^{\prime}_{2} \sigma_{2}}_{\alpha_{1} \alpha_{2}} \cdots W^{L, \sigma^{\prime}_{L} \sigma_{L}}_{\alpha_{L-1} \alpha_{0}}.
\end{equation}
The summation of $\alpha_{i}$ runs from $1$ to $\chi$, the bond dimension of the MPO. $\sigma^{\prime}_{i}$ and $\sigma_{i}$ are the indices of a local physical basis and are the same as for the MPS. The local tensors $W^{i, \sigma^{\prime}_{i} \sigma_{i}}_{\alpha_{i-1} \alpha_{i}}$ represent the operator (here $H$) but are not unique. Often it is possible to compressed them into another MPO with smaller bond dimension. The coefficient $H^{\sigma^{\prime}_{1} \cdots \sigma^{\prime}_{L}}_{\sigma_{1} \cdots \sigma_{L}}$ for open boundary condition have the diagrammatic notation
\begin{equation}
  \begin{tikzpicture}[every node/.style={scale=1},scale=0.5]
    \draw (-12.5, 0) node {$H^{\sigma^{\prime}_{1} \cdots \sigma^{\prime}_{L}}_{\sigma_{1} \cdots \sigma_{L}}$} (-10.65, 0) node {$=$};
    \draw (-9.5,1.25) node[above] {$\sigma^{\prime}_1$} (-9.5,-1.25) node[below] {$\sigma_1$} (-9.5,-2)node[below]{$W^{1}$};
    \draw (-9.5,-1.25)--(-9.5,-0.5) (-9.5,0.5)--(-9.5,1.25) (-9.5,0)circle (0.5) (-9,0)--(-7.5,0);
    \draw (-7,1.25) node[above] {$\sigma^{\prime}_2$} (-7,-1.25) node[below] {$\sigma_2$}(-7,-2)node[below]{$W^{2}$};
    \draw (-7,-1.25)--(-7,-0.5) (-7,0.5)--(-7,1.25) (-7,0)circle(0.5) (-6.5,0)--(-5,0);
    \draw (-4.5,-1.25)--(-4.5,-0.5) (-4.5,0.5)--(-4.5,1.25) (-4.5,0)circle(0.5) (-4,0)--(-3,0);
    \draw (-4.5,1.25) node[above] {$\sigma^{\prime}_3$} (-4.5,-1.25)node[below]{$\sigma_3$} (-4.5,-2)node[below]{$W^{3}$};
    \draw (-2.15,0) node {$\cdots$} (-1.5,0)--(-0.5,0);
    \draw (0,1.25) node[above] {$\sigma^{\prime}_i$} (0,-1.25)node[below]{$\sigma_i$} (0,-2)node[below]{$W^{i}$};
    \draw (0,-1.25)--(0,-0.5) (0,0.5)--(0,1.25) (0,0)circle(0.5) (0.5,0)--(1.5,0);
    \draw (2.35,0) node {$\cdots$} (3,0)--(4,0);
    \draw (4.5,1.25) node[above] {$\sigma^{\prime}_{L}$} (4.5,-1.25) node[below]{$\sigma_{L}$} (4.5,-2)node[below]{$W^{L}$};
    \draw (4.5,-1.25)--(4.5,-0.5) (4.5,0.5)--(4.5,1.25) (4.5,0)circle(0.5) (5.5,-0.25)node{$,$};
  \end{tikzpicture}
  \label{Eq:MPOCoef}
\end{equation}
with the circles representing the local tensors. The open and connected legs follows the same rules as in MPS.

\subsection{Canonical form of an MPS}\label{app:canonicalform}

The canonical form is a special gauge for TNS with loop-free structure~\cite{xiang2023density}. For MPS with open boundary conditions, by using a sequential SVD, the coefficient in canonical form can be represented as
\begin{equation}
    \psi^{\sigma_{1} \cdots \sigma_{L}} = \sum_{\{\sigma_{i}, \alpha_{i}\}} U^{1, \sigma_{1}}_{\alpha_{0} \alpha_{1}} \cdots U^{i, \sigma_{i}}_{\alpha_{i-1} \alpha_{i}} C^{i}_{\alpha_{i},\alpha_{i}} V^{i+1, \sigma_{i+1}}_{\alpha_{i} \alpha_{i+1}} \cdots  V^{L, \sigma_{L}}_{\alpha_{L-1} \alpha_{0}},
    \label{Eq:canonical_form}
\end{equation}
or graphically as
\begin{equation}
\centering
  \begin{tikzpicture}[every node/.style={scale=1},scale=0.5]
    \draw (-11.25, 0) node {$\psi^{\sigma_{1} \cdots \sigma_{L}}$} (-9.25, 0) node {$=$} (-8.75,0)--(-8,0);
    \draw (-7.5,0.5)--(-7.5,1.25) (-8,0.5)--(-7.25,0.5)--(-7,0)--(-7.25,-0.5)--(-8,-0.5)--(-8,0.5) (-7,0)--(-6,0);
    \draw (-7.5,1.25) node[above] {$\sigma_{1}$} (-7.5,-0.5)node[below] {$U^{1}$};
    \draw (-5.15,0) node{$\cdots$} (-4.5,0)--(-3.5,0);
    \draw (-3,0.5)--(-3,1.25) (-3.5,0.5)--(-2.75,0.5)--(-2.5,0)--(-2.75,-0.5)--(-3.5,-0.5)--(-3.5,0.5) (-2.5,0)--(-1.5,0);
    \draw (-3,1.25) node[above] {$\sigma_{i}$} (-3,-0.5)node[below] {$U^{i}$};
    \draw (-1,0)circle(0.5) (-1,-0.5)node[below]{$C^{i}$} (-0.5,0)--(0.5,0);
    \draw (1,0.5)--(1,1.25) (0.5,0)--(0.75,0.5)--(1.5,0.5)--(1.5,-0.5)--(0.75,-0.5)--(0.5,0) (1.5,0)--(2.5,0);
    \draw (1,1.25) node[above] {$\sigma_{i+1}$} (1,-0.5)node[below] {$V^{i+1}$};
    \draw (3.35,0) node{$\cdots$} (4,0)--(5,0);
    \draw (5.5,0.5)--(5.5,1.25) (5,0)--(5.25,0.5)--(6,0.5)--(6,-0.5)--(5.25,-0.5)--(5,0) (6,0)--(6.75,0);
    \draw (5.5,1.25) node[above] {$\sigma_{L}$} (5.5,-0.5)node[below] {$V^{L}$}(7.25,-0.25)node{$.$};
    \end{tikzpicture}
\label{eq:MPSCoefCano}
\end{equation}
where the isometric tensors $U^{i}$ and $V^{j}$ are left- and right-canonicalized, respectively. This means they satisfy the corresponding canonical conditions 
\begin{equation}
    \sum_{\sigma_{i},\alpha_{i-1}} U^{i,\sigma_{i}\raisebox{0.25ex}{*}}_{\alpha_{i-1},\alpha^{\prime}_{i}} U^{i,\sigma_{i}}_{\alpha_{i-1},\alpha_{i}} = \delta_{\alpha^{\prime}_{i},\alpha_{i}},\ \ \ 
    \sum_{\sigma_{j},\alpha_{j+1}} V^{j,\sigma_{j}\raisebox{0.25ex}{*}}_{\alpha^{\prime}_{j},\alpha_{j+1}} V^{j,\sigma_{j}}_{\alpha_{j},\alpha_{j+1}} = \delta_{\alpha^{\prime}_{j},\alpha_{j}},
\end{equation}
with the graphical representation
\begin{equation}
\centering
  \begin{tikzpicture}[every node/.style={scale=1},scale=0.5]
  \draw [rounded corners] (-10,1)--(-11,1)--(-11,-1)--(-10,-1);
  \draw (-9.5,0.5)--(-9.5,-0.5) (-10,1.5)--(-9.25,1.5)--(-9,1)--(-9.25,0.5)--(-10,0.5)--(-10,1.5);
  \draw (-10,-1.5)--(-9.25,-1.5)--(-9,-1)--(-9.25,-0.5)--(-10,-0.5)--(-10,-1.5);
  \draw (-9,1)--(-8.25,1) (-9,-1)--(-8.25,-1);
  \draw (-7.5,0) node{$=$};
  \draw [rounded corners] (-6,1)--(-6.75,1)--(-6.75,-1)--(-6,-1);
  \draw (-4.5,-0.25) node{$\quad,\quad$};
  \begin{scope}[shift={(2,0)}]
  \draw (-4,1)--(-4.75,1) (-4,-1)--(-4.75,-1);
  \draw (-3.5,-0.5)--(-3.5,0.5) (-4,1)--(-3.75,1.5)--(-3,1.5)--(-3,0.5)--(-3.75,0.5)--(-4,1);
  \draw (-4,-1)--(-3.75,-1.5)--(-3,-1.5)--(-3,-0.5)--(-3.75,-0.5)--(-4,-1);
  \draw [rounded corners](-3,1)--(-2,1)--(-2,-1)--(-3,-1);
  \draw (-1.25,0) node{$=$};
  \draw [rounded corners] (-0.5,1)--(0.25,1)--(0.25,-1)--(-0.5,-1);
  \draw (0.75,-0.25) node{$.$};
  \end{scope}
  \end{tikzpicture}
\label{eq:CanoCond}
\end{equation}
The matrix $C^{i}$ is a diagonal bond matrix obtained from the singular values $s^{i}_{m}$ of the SVD, encoding the entanglement between the left $i$ local tensors and the remaining tensors.

\section{Layout scheme}\label{app:layoutscheme}

\subsection{Layout schemes of the bosonic matrix models with two matrices}

In what follows, we will mainly focus the discussion on the layout schemes for the bosonic matrix models with two matrices.

\begin{figure}[htbp!]
    \centering
    \includegraphics[width=0.95\linewidth]{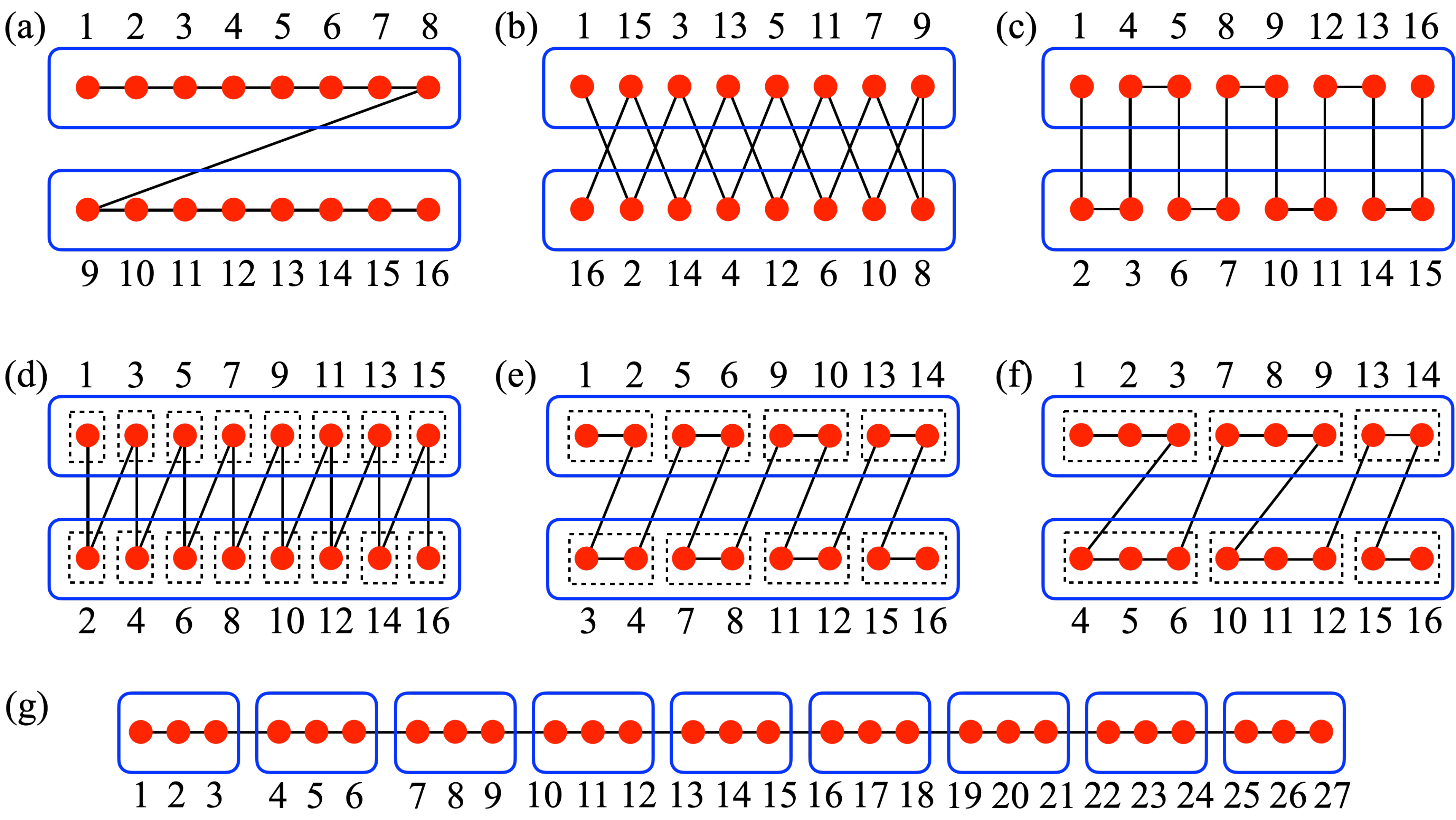}
    \caption{The schematic figures illustrating one-dimensional chain layout schemes of bosonic matrix models introduced in Sec.~\ref{sec:toyModel}. Here, each blue box represents a matrix, the red-filled circles denote the bosonic degrees of freedom $(i,\alpha)$ within each matrix $i$. The numbers in each figure represents the order of the different bosons in the MPS. The schematic figures of sequential layout scheme (a), zigzag layout scheme (b), snake path layout scheme (c), parallel layout scheme with one-site unit cell (d), with two-site unit cell (e), and with three-site unit cell (f) for the two bosonic SU$(3)$ matrices. The dashed line boxes in the parallel layout scheme denote the unit cell. The sequential layout scheme (g) for the bosonic matrix model with nine SU$(2)$ matrices.}
    \label{fig:layout_BMM_setup}
\end{figure}

As a starting point, we present some layout schemes of the SU$(N)$ bosonic matrix models with two matrices explored in this paper. The SU$(3)$ cases have been chosen to serve as a representative examples to explain the layout schemes without loss of generality. For the two matrices case, we have tested the layout schemes including the sequential layout scheme shown in Fig.~\ref{fig:layout_BMM_setup}(a), the zigzag layout scheme shown in Fig.~\ref{fig:layout_BMM_setup}(b), the snake path layout scheme shown in Fig.~\ref{fig:layout_BMM_setup}(c), and parallel layout schemes shown in Fig.~\ref{fig:layout_BMM_setup}(d), Fig.~\ref{fig:layout_BMM_setup}(e) and Fig.~\ref{fig:layout_BMM_setup}(f). Here, the blue box represents a matrix while the red-filled dots denote the bosons within each matrix. In each figure, we attach the location number of the bosons as well as a black line to explain the ordering patterns in different layout schemes. To explain the parallel layout schemes, we introduce a dashed line box to represent a \textit{unit cell} in Fig.~\ref{fig:layout_BMM_setup}(d-f). Note that, as in Fig.~\ref{fig:layout_BMM_setup}(f), the last unit cell might be smaller than the others when the size of the matrices is not a multiple of the unit cell's size. These parallel layout schemes with unit cells of size one-, two-, and three-site can be generalized straight forwardly.

The next step is to estimate the maximal bond dimension of the Hamiltonian MPO for all layout schemes shown in Fig.~\ref{fig:layout_BMM_setup}(a-f). Instead of using operator entanglement~\cite{prosen2007operator,vznidarivc2008complexity} as an indicator, we chose the AutoMPO function in ITensors~\cite{fishman2022itensor} to directly compress MPO by SVD. During the compression process, the truncation error is set as $10^{-15}$ to ensure the sufficient precision when compressing, and this value is adopted for all subsequent computations throughout this paper. Table~\ref{tab:DMPO_2BMM_noPenalty} presents the maximal bond dimension of the Hamiltonian MPO obtained from AutoMPO for two SU$(N)$ bosonic matrices without the penalty terms. The red bold number indicates the minimal bond dimension among all layout schemes, and the corresponding layout scheme for each SU$(N)$ is adopted in our computation in the two bosonic matrices cases.

\begin{table}[h!]
    \centering
    \caption{The maximal bond dimension of the Hamiltonian MPO for two SU$(N)$ bosonic matrices without the penalty terms.}
    \begin{tabular}{c|c|c|c|c|c}
        \hline
        \hline
           & $N$ = 2 & $N$ = 3 & $N$ = 4 & $N$ = 5 & $N$ = 6 \\
        \hline
        Sequential layout scheme & 8 & 38 & 122 & 302 & 632 \\
        \hline
        Zigzag layout scheme & \textcolor{red}{\textbf{7}} & 49 & 131 & 290 & 570 \\
        \hline
        Snake path layout scheme & 8 & 37 & 86 & 140 & 206 \\
        \hline
        \makecell{Parallel layout scheme\\ with single-site unit cell} & 8 & 37 & 86 & 140 & 206 \\
        \hline
        \makecell{Parallel layout scheme\\ with two-site unit cell} & 8 & 37 & 77 & 140 & 189 \\
        \hline
        \makecell{Parallel layout scheme\\ with three-site unit cell} & 8 & \textcolor{red}{\textbf{36}} & 86 & \textcolor{red}{\textbf{126}} & \textcolor{red}{\textbf{188}} \\
        \hline
        \makecell{Parallel layout scheme\\ with four-site unit cell} & 8 & 37 & \textcolor{red}{\textbf{76}} & 140 & 188 \\
        \hline
        \makecell{Parallel layout scheme\\ with five-site unit cell} & 8 & 37 & 77 & 126 & 206 \\
        \hline
        \makecell{Parallel layout scheme\\ with six-site unit cell} & 8 & 42 & 83 & 132 & 188 \\
        \hline
        \makecell{Parallel layout scheme\\ with seven-site unit cell} & 8 & 46 & 77 & 127 & 194 \\
        \hline
        \makecell{Parallel layout scheme\\ with eight-site unit cell} & 8 & 38 & 76 & 148 & 188 \\
        \hline
        \hline
        \end{tabular}
        \label{tab:DMPO_2BMM_noPenalty}
\end{table}

\subsection{Layout scheme of the bosonic matrix models with more than two matrices}

We now show some layout scheme for bosonic models with more than two matrices. In this paper, we focus on the SU$(2)$ and SU$(3)$ bosonic matrix models when the number of matrices exceeds two. As shown in Tab.~\ref{tab:DMPO_2BMM_noPenalty}, the Hamiltonian MPO in these cases have nearly the same maximal bond dimension. Therefore, we directly adopt the sequential layout scheme for convenience, rather than conducting additional benchmarks to select the optimal layout scheme as we did for the two-matrix cases. Fig.~\ref{fig:layout_BMM_setup}(g) shows the sequential layout scheme for nine bosonic matrices. Following the same convention as in the case of two matrices, the blue box and red dots represent each a matrix and a bosonic degree of freedom, respectively. The solid line, as well as the number, sever as the reference to indicate the labelling order of the bosons in the MPS.

Table~\ref{tab:DMPO_9BMM_noPenalty} gives the maximal bond dimension of SU$(2)$ and SU$(3)$ bosonic matrix models, both with and without the gauge-invariant penalty term $p G^2$. For both the SU$(2)$ and SU$(3)$ case, the maximal bond dimension with the gauge invariant penalty is larger than the cases without gauge. This additional bond dimension arises because more terms related to the penalty term are included in the Hamiltonian. Furthermore, as we move from the SU$(2)$ to the SU$(3)$ case, the maximal bond dimension increases significantly. However, in all the four cases, the maximal bond dimension begins to saturate when the number of matrices exceeds two. 

\begin{table}[h!]
    \centering
    \caption{Bond dimension of the MPO of SU$(2)$ bosonic matrices without and with the gauge invariant penalty term using the layout scheme $1$}
    \begin{tabular}{c|c|c}
        \hline
        \hline
         & Two matrices & More than two matrices \\
        \hline
         SU(2) without penalty terms & 8 & 12 \\
        \hline
         SU(2) with penalty terms & 13 & 21 \\
        \hline
         SU(3) without penalty terms &  38 & 50 \\
         \hline
         SU(3) with penalty terms & 56 & 81 \\
         \hline
         \hline
    \end{tabular}
    \label{tab:DMPO_9BMM_noPenalty}
\end{table}

\subsection{Layout scheme for the minimal BMN-like model}

\begin{figure}[htbp!]
    \centering
    \includegraphics[width=0.9\linewidth]{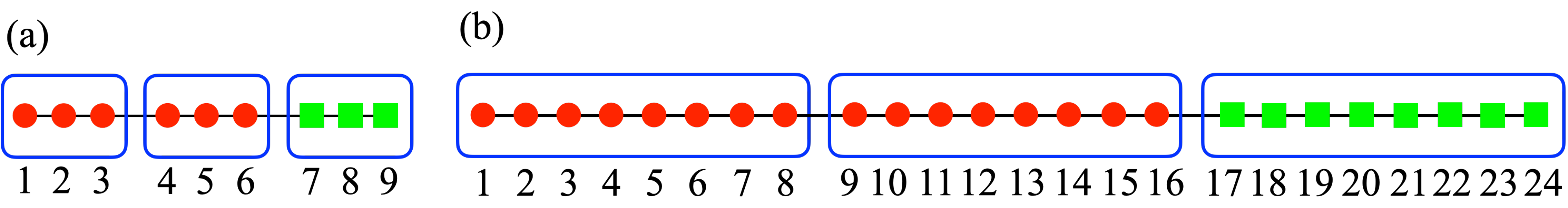}
    \caption{The schematic figures of sequential layout scheme illustrate one-dimensional chain layout schemes of the minimal (a) SU$(2)$ and (b) SU$(3)$ BMN models with two matrices introduced in Sec.~\ref{sec:toyModel}. The blue boxes represent matrices, the red-filled circles and green squares denote the bosons and fermions respectively. The numbers represent the sequence of the different bosons and fermions, all of which are connected by the black solid line along the chain to emphasize the order.}
    \label{fig:layout_BMN_setup}
\end{figure}

The minimal BMN-like model Eq.~\eqref{eq:toymodel} consists of two bosonic matrices and the corresponding fermion matrix. Since our aim is to demonstrate the feasibility of using TNS to study the BMN model, we will only focus on the sequential layout scheme as an example, without further exploration of alternative schemes. As fig.~\ref{fig:layout_BMN_setup} shows, we present the sketches of the SU$(2)$ and SU$(3)$ minimal BMN models. Following the conventions used in the matrix models above, the blue boxes filed with red circles indicating boson matrices. The green filled squares denote the accompanying fermions, positioned after all bosonic sites. For clarity, the serial numbers as well as the black lines are included to explain this sequential layout scheme. The Hamiltonian MPO without the gauge-invariant penalty terms can be brought down to bond dimension \textbf{11} and \textbf{46} for SU$(2)$ and SU$(3)$, respectively. They will increase to \textbf{20} and \textbf{64} when penalty terms are included. 

\bibliographystyle{JHEP}
\bibliography{lit}

\end{document}